\documentclass[11pt,a4paper]{article}
\usepackage{amsmath}
\usepackage{amsfonts}
\usepackage{ulem}
\usepackage{color}
\usepackage[thicklines]{cancel}
\usepackage{graphicx}
\usepackage{dcolumn}
\usepackage{bm}
\usepackage[
bookmarks,bookmarksnumbered,colorlinks=true,anchorcolor=blue,
linkcolor=blue,urlcolor=blue,citecolor=blue,
breaklinks=true]{hyperref}
\usepackage{geometry}
\usepackage{authblk}
\usepackage{amssymb}
\usepackage[utf8]{inputenc}
\usepackage[numbers,sort&compress]{natbib}
\usepackage{pifont}
\usepackage{tikz}
\usepackage{ulem}

\numberwithin{equation}{section}   

\date{\today}

\begin{document}

\title{\bf Nearly critical superfluid: \\ effective field theory and holography}

\author[2]{Yanyan Bu \thanks{yybu@hit.edu.cn}}
\author[1]{Hongfei Gao \thanks{gaohf1999@gmail.com (co-correspondence author)}}
\author[1]{Xin Gao \thanks{xingao@scu.edu.cn (co-correspondence author)}}
\author[1]{Zhiwei Li \thanks{lizhiwei3@stu.scu.edu.cn (co-correspondence author)}}

\affil[2]{\it School of Physics, Harbin Institute of Technology, Harbin 150001, China} 
\affil[1]{\it College of Physics, Sichuan University, Chengdu 610065, China}

\maketitle

\begin{abstract}
We study a nearly critical superfluid system from two complementary approaches. Within the first approach, we formulate a Schwinger-Keldysh effective field theory (EFT) for the system when it is located slightly above the critical temperature. The dynamical variables in the EFT construction are two scalars: a neutral scalar associated with the conserved U(1) charge, and a complex scalar describing the order parameter. The set of symmetries, particularly the dynamical Kubo-Martin-Schwinger (KMS) symmetry and chemical shift symmetry, strictly constrains the form of EFT action. Within the second approach, using the holographic Schwinger-Keldysh technique, we derive the effective action for a ``microscopic'' holographic superfluid, confirming the EFT construction. A systematic inclusion of non-Gaussianity is one highlight of present study.
\end{abstract}

\newpage

\tableofcontents

\allowdisplaybreaks

\flushbottom

\section{Introduction}

Non-equilibrium phenomena are ubiquitous in nature. However, in contrast to equilibrium situation, we still lack a unified framework for understanding diverse non-equilibrium phenomena. One practical approach is to model specific cases. Critical phenomena and phase transitions have been an important research subject, pushing forward development of non-equilibrium statistical physics. Modern theory of (dynamical) critical phenomena takes the fact that systems that look quite different microscopically could share identical critical exponents and thus belong to the same ``universality'' class \cite{RevModPhys.49.435}. This idea motivated physicists to build an effective model for each universality class, which takes the form of a stochastic partial differential equation (PDE). The stochasticity is due to fluctuation-dissipation theorem and is usually realized by a random variable obeying Gaussian distribution, mimicking thermal fluctuation. Moreover, stochastic models could be cast into Martin-Siggia-Rose (MSR) formalism, allowing one to study non-equilibrium systems using standard field theoretic techniques \cite{kamenev_2023,tauber_2014}. Nowadays, this framework becomes an indispensable tool in exploring non-equilibrium dynamics.

The effectiveness of stochastic models in the study of critical phenomena could be understood from the perspective of Wilsonian renormalization group (RG). While systems from a same universality class may show remarkable differences at microscopic scale, they will flow to the same infrared (IR) fixed point (the critical point), and thus share the same effective description emergent in the critical regime.

By virtue of Wilsonian RG, an effective field theory (EFT) has been recently formulated for dissipative hydrodynamics \cite{Crossley:2015evo,Glorioso:2017fpd,Haehl:2015uoc,Haehl:2018lcu}\footnote{For early attempts on this subject, see e.g., \cite{Dubovsky:2011sj,Endlich:2012vt,Grozdanov:2013dba,
Kovtun:2014hpa,Haehl:2015foa,Montenegro:2016gjq}. Further exploration on formal aspects of hydrodynamic EFT can be found in e.g., \cite{Glorioso:2016gsa,Gao:2018bxz,Jensen:2017kzi}.} (see \cite{Liu:2018kfw} for a nice review). The hydrodynamic EFT cares about dynamics of conserved quantities (such as energy, momentum, internal charge), which are the only dynamical modes surviving in the hydrodynamic regime. In order to capture both fluctuation and dissipation, hydrodynamic EFT is formulated using the Schwinger-Keldysh formalism, in which the degrees of freedom are doubled. In addition, a set of symmetries is proposed to constrain the hydrodynamic EFT action. Notably, in contrast to stochastic models, hydrodynamic EFT provides a systematic treatment of fluctuations and dissipations at full nonlinear level. Therefore, the methodology of hydrodynamic EFT becomes an ideal framework for investigating fluctuation effects. Indeed, over the past few years, such a new methodology has attracted a lot of attention in diverse physical settings, see e.g., \cite{Blake:2017ris,Chen-Lin:2018kfl,Hongo:2019qhi,Jain:2020zhu,Baggioli:2020haa,
Landry:2020ire,Sogabe:2021svv,Vardhan:2022wxz,Landry:2022nog,
Abbasi:2022aao,Lin:2023bli,Donos:2023ibv,Gao:2023wun,Glorioso:2023chm,Jain:2023obu,
Mullins:2023ott,Delacretaz:2023ypv,Delacretaz:2023pxm,Huang:2023eyz}.

Holographic duality \cite{Maldacena:1997re,Gubser:1998bc,Witten:1998qj} has been insightful in the study of non-equilibrium physics, particularly in the development of hydrodynamic EFT. Firstly, holography has enlightened the choosing of suitable dynamical variables for writing hydrodynamic EFT \cite{Nickel:2010pr,Crossley:2015tka}. Secondly, holographic Schwinger-Keldysh technique \cite{Glorioso:2018mmw} (see \cite{Herzog:2002pc,Skenderis:2008dh,Skenderis:2008dg} for alternative approaches) provides a tractable tool for {\it deriving} effective action for a certain holographic model \cite{Glorioso:2018mmw,deBoer:2018qqm,Chakrabarty:2019aeu,Bu:2020jfo,Bu:2021clf,
Bu:2021jlp,Bu:2022esd,Bu:2022oty,Baggioli:2023tlc}\footnote{Similar study was carried out in \cite{Ho:2013rra,Ghosh:2020lel,He:2021jna}. We understand that it is the influence functional rather than the off-shell effective action that was obtained therein.}. The latter point is important on its own right: holographic study will contribute to examining various symmetry postulates in the construction of hydrodynamic EFT, and may even shed light on generalization of current framework of hydrodynamic EFT.

When extra modes (apart of those conserved quantities) happen to be relevant\footnote{This corresponds to quasi hydrodynamics \cite{Grozdanov:2018fic}, in which the strict hydrodynamics is enlarged in order to cover a mode with a small gap.}, the framework of hydrodynamic EFT shall be enlarged. The critical dynamics near a phase transition offers such an example. The phenomenon of critical slowing down indicates that the order parameter shall be retained in the low energy EFT. It is then interesting to formulate an EFT for critical dynamics near a phase transition, which has been recently attacked in the context of a nearly critical superfluid system in \cite{Bu:2021clf,Lin:2023bli,Donos:2023ibv}. Ref. \cite{Bu:2021clf} employed the holographic Schwinger-Keldysh technique of \cite{Glorioso:2018mmw}, and focused on dynamics of a fluctuating order parameter, particularly on a systematic inclusion of non-Gaussian noises. Meanwhile, the charge diffusion sector was turned off for simplicity. Later on, based on the result of \cite{Bu:2021clf}, Ref. \cite{Lin:2023bli} revealed a systematic way of including non-Gaussian noises in stochastic formalisms. The work \cite{Donos:2023ibv} presented an EFT construction for a nearly critical superfluid. With various approximations undertaken, the EFT action of \cite{Donos:2023ibv} is essentially identical to the MSR formalism of Model F under the classification of Hohenberg and Halperin \cite{RevModPhys.49.435}.

In this work we will continue the study of a nearly critical superfluid system using methodology of hydrodynamic EFT \cite{Crossley:2015evo,Glorioso:2017fpd} and holographic technique \cite{Glorioso:2018mmw}. The main objective will be twofold. On the one hand, we will present a more general EFT action by relaxing various approximations assumed in previous studies \cite{Bu:2021clf,Donos:2023ibv}. On the other hand, through a direct calculation within a critical holographic superfluid model \cite{Herzog:2010vz}\footnote{Via holography, relaxation dynamics near critical regime has been recently considered in e.g., \cite{Chen:2018msc,Flory:2022uzp,Donos:2022qao,Cai:2022omk,Cao:2022csq,Zhao:2023gur}.}, we will not only confirm the general EFT construction but also provide holographic lesson for various coefficients in the effective action.

The rest of this work will be structured as follows. In section \ref{EFT_construction} we present a thorough construction for the EFT of a nearly critical superfluid system. In this section, we also comment on the relationship between present work and relevant studies in the literature. In section \ref{EFT_holo} we carry out a holographic derivation of the EFT action. First, we outline a holographic program towards boundary EFT. Second, we explain boundary conditions for bulk fields and their relationship with some symmetries used for formulating the EFT. Last, we set up a perturbation theory in the bulk, and derive boundary EFT Lagrangian. In section \ref{Summary_Outlook} we present a brief summary and outlook several future directions. In appendices \ref{source_terms} and \ref{holo_calculation}, we supplement further details regarding holographic study.

\section{Effective field theory for a nearly critical superfluid} \label{EFT_construction}

In this section we present the construction of EFT for a superfluid system near the critical point. For simplicity, throughout this work, we will not consider the dynamics of energy and momentum. Moreover, we assume that the system has been tuned to be slightly above the critical temperature. Thus, the global U(1) symmetry associated with the superfluidity is not spontaneously broken. The dynamical degrees of freedom for such a critical system are a conserved U(1) charge and a non-conserved order parameter. The non-conserved order parameter can be simply described by a complex scalar field $O_s$ (and the complex conjugate $O^*_s$ as well), with $s =1$ ($s=2$) denoting the upper (lower) branch of the SK contour. In order to write the EFT, the conserved U(1) charge is suitably described by the following gauge invariant object \cite{Crossley:2015evo}
\begin{align}
B_{s\mu} \equiv \mathcal A_{s\mu} + \partial_\mu \varphi_s, \qquad s=1 ~ {\rm or} ~2  \label{block1}
\end{align}
where $\mathcal A_{s\mu}$ is an external gauge potential, and $\varphi_s$ is the dynamical field. Indeed, instead of $O_s$ and $O^*_s$, we will find it more convenient to work with the following variables
\begin{align}
\Delta_s \equiv e^{{\rm i} q \varphi_s} O_s, \qquad \qquad  \Delta_s^* \equiv e^{-{\rm i} q \varphi_s} O_s^*  \label{block2}
\end{align}
which is also motivated by holographic study in section \ref{EFT_holo}. From here on, the charge $q$ will be set to unity. The EFT action is a local functional of the building blocks $B_{s\mu}$, $\Delta_s$ and $\Delta_s^*$
\begin{align}
S_{eff} = S_{eff}[B_{1\mu}, \Delta_1, \Delta_1^*; B_{2\mu}, \Delta_2, \Delta_2^*] = S_{eff}[B_{r\mu}, \Delta_r, \Delta_r^*; B_{a\mu}, \Delta_a, \Delta_a^*],
\end{align}
Here, the Keldysh basis is defined as
\begin{align}
B_{r\mu} \equiv \frac{1}{2}(B_{1\mu} + B_{2\mu}), \qquad \qquad B_{a\mu} \equiv B_{1\mu} - B_{2\mu}
\end{align}
and similarly for $\Delta_{r,a}$ and $\Delta_{r,a}^*$. Based on the EFT action, the partition function of the critical superfluid system is
\begin{align}
Z[\mathcal A_{1\mu}; \mathcal A_{2\mu}] = \int [D \varphi_s] [D \Delta_s] [D \Delta_s^*]\, e^{{\rm i} S_{eff}[B_{1\mu}, \Delta_1, \Delta_1^*; B_{2\mu}, \Delta_2, \Delta_2^*]} \label{Z_EFT}
\end{align}

\subsection{The full set of symmetries}

Here we list out all the symmetries that the EFT action $S_{eff}$ shall satisfy.

$\bullet$ Normalization condition. This condition requires the EFT action to be vanishing if the fields living on two SK legs are set identical
\begin{align}
S_{eff} [B_{1\mu}, \Delta_1, \Delta_1^*; B_{1\mu}, \Delta_1, \Delta_1^*] = 0
\end{align}
which implies the EFT action must contain at least one factor of $a$-variable. Indeed, the normalization condition is motivated by the following property of the partition function
\begin{align}
Z[\mathcal A_{1\mu} = \mathcal A_{2\mu}]=1, \label{normalization_Z}
\end{align}
which will be apparent if one rewrites \eqref{Z_EFT} as a path integral over microscopic degrees of freedom \cite{Crossley:2015evo}. In order to ensure \eqref{normalization_Z} at the level of full path integral \eqref{Z_EFT}, the normalization condition shall be enlarged into a BRST-type symmetry. Here, one introduces a ghost partner for each of the dynamical fields, and add a ghost action to the original action $S_{eff}$ so that the total action is invariant under a BRST-type transformation. We recommend the interesting readers to \cite{Haehl:2015foa,Crossley:2015evo,Jensen:2017kzi} for more details.

$\bullet$ $Z_2$ reflection symmetry
\begin{align}
    S_{eff}^* [B_{1\mu}, \Delta_1, \Delta_1^*; B_{2\mu}, \Delta_2, \Delta_2^*] = - S_{eff} [B_{2\mu}, \Delta_2, \Delta_2^*; B_{1\mu}, \Delta_1, \Delta_1^*],
\end{align}
which implies the action $S_{eff}$ must contain complex coefficients. Obviously, this $Z_2$ reflection symmetry ensures that $Z^*[\mathcal A_{1\mu}; \mathcal A_{2\mu}] = Z[\mathcal A_{2\mu}; \mathcal A_{1\mu}] $, which is a basic property of the SK formalism \cite{Crossley:2015evo}.

Indeed, both normalization condition and $Z_2$ reflection symmetry are related to unitarity of time evolution.

$\bullet$ Imaginary part of $S_{eff}$ is positive-definite
\begin{align}
{\rm Im}(S_{eff}) \geq 0 \label{Z_well-defined}
\end{align}
so that the path integral based on EFT action, cf. \eqref{Z_EFT}, is well-defined.

$\bullet$ Spatially rotational symmetry. This guides one to classify building blocks and their derivatives according to SO(3) spatially rotational transformation.

$\bullet$ Global U(1) symmetry. This symmetry governs the coupling between the conserved U(1) charge and the complex order parameter. Recall that the present work focuses on the high temperature phase so that the global U(1) symmetry is unbroken. So, the effective action $S_{eff}$ is invariant under a {\it diagonal} global U(1) transformation. This is automatically guaranteed if the variables $\Delta_s$ and $\Delta_s^*$ will appear simultaneously in the action $S_{eff}$.

$\bullet$ Chemical shift symmetry. This is due to the fact that the global U(1) symmetry is not broken spontaneously. This symmetry will act on the diffusive fields $\varphi_s$, and amounts to defining what we mean by a normal diffusion. More precisely, the EFT action $S_{eff}$ is invariant under the following diagonal time-independent shift over $\varphi_s$
\begin{align}
    \varphi_r \to \varphi_r + \sigma(\vec x), \qquad \varphi_a \to \varphi_a, \qquad {\rm others~unchanged}. \label{chemical_shift}
\end{align}
Obviously, under the shift \eqref{chemical_shift}, the building blocks \eqref{block1} and \eqref{block2} transform as
\begin{align}
& B_{ri} \to B_{ri} + \partial_i \sigma(\vec x), \qquad B_{r0} \to B_{r0}, \qquad B_{a\mu} \to B_{a\mu}, \nonumber \\
& \Delta_{r,a} \to e^{{\rm i}\sigma(\vec x)} \Delta_{r,a}, \qquad \quad \Delta_{r,a}^* \to e^{- {\rm i} \sigma(\vec x)} \Delta_{r,a}^*. \label{chemical_shift1}
\end{align}
Analogous to QED, one could introduce a covariant derivative operator $\mathcal D_i$, associated with $B_{ri}$, acting on the complex order parameter \cite{Donos:2023ibv}
\begin{align}
    \mathcal D_i \Delta_{r,a} \equiv \partial_i \Delta_{r,a} - {\rm i} B_{ri} \Delta_{r,a}, \qquad  (\mathcal D_i \Delta_{r,a})^* \equiv \partial_i \Delta_{r,a}^* + {\rm i} B_{ri} \Delta_{r,a}^* \label{covariant_der}
\end{align}
which, under the chemical shift \eqref{chemical_shift}, transform in the same fashion as $\Delta_{r,a}$ and $\Delta_{r,a}^*$
\begin{align}
\mathcal D_i \Delta_{r,a} \to e^{{\rm i} \sigma(\vec x)}\mathcal D_i \Delta_{r,a}, \qquad  (\mathcal D_i \Delta_{r,a})^* \to e^{- {\rm i} \sigma(\vec x)} (\mathcal D_i \Delta_{r,a})^*.
\end{align}
Therefore, instead of $\partial_i \Delta_{r,a}$ and $\partial_i \Delta_{r,a}^*$, we will use the covariant derivatives $\mathcal D_i \Delta_{r,a}$ and $(\mathcal D_i \Delta_{r,a})^*$ when constructing the action $S_{eff}$. Interestingly, this symmetry links terms containing different number of fields.

Given the chemical shift symmetry, $B_{ri}$ would appear in the EFT action through the following objects: $\partial_0 B_{ri}$, $\mathcal D_i \Delta_{r,a}$, $(\mathcal D_i \Delta_{r,a})^*$ or $\mathcal F_{rij} \equiv \partial_i B_{rj} - \partial_j B_{ri}$.

$\bullet$ Dynamical KMS symmetry. When the physical system is in a thermal state, the KMS condition sets important constraint on the generating functional $W = - {\rm i} \log Z$. The KMS condition can be expressed in terms of $n$-point correlation functions (i.e., functional derivatives of $W$ with respect to external sources), generalizing familiar FDT to nonlinear case \cite{Wang:1998wg,Hou:1998yc} (see also \cite{Crossley:2015evo}). Obviously, the KMS condition and the generalized nonlinear FDT are valid at the full quantum level. Within hydrodynamic EFT framework, KMS condition is guaranteed by the proposal that $S_{eff}$ shall satisfy dynamical KMS symmetry \cite{Glorioso:2016gsa,Glorioso:2017fpd}. In the classical statistical limit where quantum fluctuations are ignored, the dynamical KMS symmetry gets simplified
\begin{align}
S_{eff}[B_{r\mu}, \Delta_r, \Delta_r^*; B_{a\mu}, \Delta_a, \Delta_a^*] = S_{eff}[\widehat B_{r\mu}, \widehat \Delta_r, \widehat \Delta_r^*; \widehat B_{a\mu}, \widehat\Delta_a, \widehat \Delta_a^*], \label{KMS_symmetry}
\end{align}
where
\begin{align}
& \widehat B_{r\mu}(-v,-\vec x)= (-1)^{\eta_\mu}B_{r\mu}(v, \vec x), \qquad \widehat B_{a\mu}(-v, -\vec x) = (-1)^{\eta_\mu} \left[ B_{a\mu}(v, \vec x) + {\rm i} \beta \partial_0 B_{r\mu}(v, \vec x) \right], \nonumber \\
& \widehat \Delta_r(-v, -\vec x) = (-1)^{\eta_\Delta} \Delta_r^*(v,x), \qquad \widehat \Delta_a(-v, -\vec x) = (-1)^{\eta_\Delta}\left[ \Delta_a^*(v,x) + {\rm i} \beta \partial_0 \Delta_r^*(v,\vec x) \right], \nonumber \\
& \widehat \Delta_r^*(-v, -\vec x) = (-1)^{\eta_\Delta} \Delta_r(v,x), \qquad \widehat \Delta_a^*(-v, -\vec x) = (-1)^{\eta_\Delta}\left[ \Delta_a(v,x) + {\rm i} \beta \partial_0 \Delta_r(v,\vec x) \right]. \label{KMS_transform}
\end{align}
Here, $v$ is the time coordinate, and $\beta$ is inverse of temperature; $(-1)^{\eta_\mu} = +1$ and $(-1)^{\eta_\Delta} = -1$ are eigenvalues of $B_\mu$ and $\Delta$, respectively, under discrete symmetries $\mathcal{PT}$. This symmetry sets a link between terms with different number of time derivatives but equal number of fields.

$\bullet$ Onsager relations. This requirement follows from the symmetry properties of
the retarded (or advanced) correlation functions under a change of the ordering of operators \cite{Crossley:2015evo}. While for some simple cases, Onsager relations are satisfied automatically once dynamical KMS symmetry is imposed, this is not generically true (see \cite{Crossley:2015evo,Baggioli:2023tlc} for further examples). We will see this fact regarding some quartic terms in the action.

\subsection{EFT action}

With suitable variables and symmetries identified, it is ready to write down the effective action for the critical superfluid system. Basically, as in any EFT, we will organize the effective action by number of fields and number of spacetime derivatives. Accordingly, the effective action will be split as follows
\begin{align}
    S_{eff} = \int d^4x \mathcal L_{eff} = \int d^4x \left[ \mathcal L_{diff} + \mathcal L_{\Delta} + \mathcal L_{int} \right],
\end{align}
where $\mathcal L_{diff}$ is effective Lagrangian for the U(1) charge diffusion; $\mathcal L_{\Delta}$ is that of a complex order parameter; and $\mathcal L_{int}$ represents interactions of the diffusive field and the order parameter. We proceed to write down the effective Lagrangian by imposing some of the symmetries, i.e., normalization condition, $Z_2$ reflection symmetry, spatially rotational symmetry, global U(1) symmetry, and chemical shift symmetry. Afterwards we will come back to constraints arising from the rest symmetries.

$\bullet$ EFT Lagrangian for the diffusion $\mathcal L_{diff}$

Here, we truncate the Lagrangian to quadratic order in diffusive fields and second order in spacetime derivatives. The result is
\begin{align}
\mathcal L_{diff} & = a_0 B_{a0} B_{r0} + a_1 B_{a0} \partial_0 B_{r0} + a_2 B_{ai} \partial_0 B_{ri} + a_3 B_{ai} \partial_i B_{r0} + a_4 B_{a0} \partial_0^2 B_{r0} \nonumber \\
&+ a_5 B_{a0} \partial_i^2 B_{r0} + a_6 B_{a0} \partial_0 \partial_i B_{ri} + a_7 B_{ai} \partial_0 \partial_i B_{r0} + a_8 B_{ai} \partial_0^2 B_{ri} + a_9 \mathcal F_{aij} \mathcal F_{rij} \nonumber \\
& + {\rm i} u_0 B_{a0}^2 + {\rm i} u_1 B_{ai}^2 + {\rm i} u_2 B_{a0} \partial_i B_{ai}  + {\rm i} u_3 B_{a0} \partial_0^2 B_{a0} + {\rm i} u_4 B_{a0} \partial_i^2 B_{a0} \nonumber \\
& + {\rm i} u_5 B_{a0} \partial_0 \partial_i B_{ai} + {\rm i} u_6 B_{ai} \partial_0^2 B_{ai} + {\rm i} u_7 B_{ai} \partial_k^2 B_{ai} + {\rm i} u_8 B_{ai} \partial_i \partial_j B_{aj}, \label{L_diff}
\end{align}
where $\mathcal F_{aij} \equiv \partial_i B_{aj} - \partial_j B_{ai}$. In actual fact, this part has been intensively studied in the literature from both EFT and holographic perspectives, see \cite{Crossley:2015evo,Glorioso:2018mmw,deBoer:2018qqm,Bu:2020jfo,Baggioli:2023tlc} for more details. Due to $Z_2$ reflection symmetry, all the coefficients in \eqref{L_diff} are purely real.

We explore constraints due to the rest symmetries. The condition \eqref{Z_well-defined} requires
\begin{align}
u_0 \geq 0, \qquad u_1 \geq 0.
\end{align}
Imposing the dynamical KMS symmetry, we find
\begin{align}
a_1 = -\beta u_0, \quad a_2 = - \beta u_1, \quad  a_3 =0, \quad u_2 =0, \quad  a_6 = a_7.
\end{align}
Then, Onsager relations are satisfied automatically.

Via KMS, the $aa$-terms with second order derivatives (i.e., $u_{3-8}$-terms) shall be linked to $ra$-terms with third order derivatives that are not presented in \eqref{L_diff}.

$\bullet$ EFT Lagrangian for order parameter $\mathcal L_{\Delta}$

As in the diffusive part $\mathcal L_{diff}$, we retain terms up to quadratic order in order parameter and second order in spacetime derivatives. Then, the Lagrangian is
\begin{align}
\mathcal L_{\Delta} & = b_0 \Delta_a^* \Delta_r + b_0^* \Delta_a \Delta_r^* + b_1 \Delta_a^* \partial_0 \Delta_r + b_1^* \Delta_a \partial_0 \Delta_r^* + b_2 \Delta_a^* \partial_0^2 \Delta_r + b_2^* \Delta_a \partial_0^2 \Delta_r^* \nonumber \\
& + b_3 (\mathcal D_i \Delta_a)^* (\mathcal D_i \Delta_r) + b_3^* (\mathcal D_i \Delta_a) (\mathcal D_i \Delta_r)^* + {\rm i} v_0 \Delta_a^* \Delta_a + v_1 \Delta_a^* \partial_0 \Delta_a \nonumber \\
&+ {\rm i} v_2 \Delta_a^* \partial_0^2 \Delta_a + {\rm i} v_3 (\mathcal D_i \Delta_a)^* (\mathcal D_i \Delta_a). \label{L_Delta}
\end{align}
Here, by $Z_2$ reflection symmetry, $v_{0,1,2,3}$ are purely real, while other ones could be complex. Notice that, in order to make the chemical shift symmetry transparent, we have utilized the covariant derivative operator $\mathcal D_i$ defined in \eqref{covariant_der}. As a result, this treatment inevitably brings in interactions between $B_{s\mu}$ and $\Delta_s, \Delta_s^*$ in the above Lagrangian. Intriguingly, $B_{ri}$ is now allowed to appear explicitly, which is forbidden in \eqref{L_diff} by chemical shift symmetry.

The condition \eqref{Z_well-defined} imposes that
\begin{align}
v_0 \geq 0.
\end{align}
From the dynamical KMS symmetry, we have
\begin{align}
b_0= b_0^*, \quad  b_1 + b_1^* = - \beta v_0, \quad  b_2 - b_2^* = i\beta v_1, \quad  b_3 = b_3^*.
\end{align}
Then, Onsager relations are satisfied automatically. Interestingly, $b_1$ and $b_2$ could be complex, which is also supported by holographic study.

$\bullet$ EFT Lagrangian for the interaction $\mathcal L_{int}$

For this part, we will keep terms to quartic order in dynamical fields, and to first order in spatial derivatives, but ignore time derivative terms. This is partially motivated by the scaling argument $\partial_0 \sim \partial_i^2$. We organize the Lagrangian by number of $a$-fields:
\begin{align}
\mathcal L_{int} & = c_0 B_{a0} \Delta_r^* \Delta_r + c_1 B_{r0} \Delta_a^* \Delta_r + c_1^* B_{r0} \Delta_a \Delta_r^* + {\rm i} c_2 B_{ai} (\mathcal D_i \Delta_r)^* \Delta_r \nonumber \\
& - {\rm i} c_2^* B_{ai} (\mathcal D_i \Delta_r) \Delta_r^* + c_3 \Delta_a \Delta_r^* \Delta_r^* \Delta_r + c_3^* \Delta_a^* \Delta_r \Delta_r^* \Delta_r + c_4 B_{r0}^2 \Delta_a^* \Delta_r \nonumber \\
& + c_4^* B_{r0}^2 \Delta_a \Delta_r^* + c_5 B_{a0} B_{r0} \Delta_r^* \Delta_r + {\rm i} w_0 B_{a0} \Delta_a^* \Delta_r + {\rm i} w_0 B_{a0} \Delta_a \Delta_r^*\nonumber \\
&+ {\rm i} w_1 B_{r0} \Delta_a^* \Delta_a  + w_2 B_{ai} (\mathcal D_i \Delta_a)^* \Delta_r - w_2^* B_{ai} (\mathcal D_i \Delta_a) \Delta_r^* + w_3 B_{ai} (\mathcal D_i \Delta_r)^* \Delta_a \nonumber \\
& - w_3^* B_{ai} (\mathcal D_i \Delta_r) \Delta_a^* + {\rm i} w_4 \Delta_a \Delta_a \Delta_r^* \Delta_r^* + {\rm i} w_4 \Delta_a^* \Delta_a^* \Delta_r \Delta_r + {\rm i} w_5 \Delta_a \Delta_a^* \Delta_r^* \Delta_r \nonumber \\
& + {\rm i} w_6 B_{ai}^2 \Delta_r^* \Delta_r + {\rm i} w_7 B_{r0}^2 \Delta_a^* \Delta_a + {\rm i} w_8 B_{a0} B_{r0} \Delta_a^* \Delta_r + {\rm i} w_8 B_{a0} B_{r0} \Delta_a \Delta_r^* \nonumber \\
& + {\rm i} w_9 B_{a0}^2 \Delta_r^* \Delta_r + w_{10} B_{a0} \Delta_a^* \Delta_a + {\rm i} w_{11} B_{ai} (\mathcal D_i \Delta_a)^* \Delta_a -{\rm i} w_{11} B_{ai} (\mathcal D_i \Delta_a) \Delta_a^* \nonumber \\
& + w_{12} \Delta_a^* \Delta_a \Delta_a \Delta_r^* + w_{12}^* \Delta_a^* \Delta_a^* \Delta_a \Delta_r + w_{13} B_{ai}^2 \Delta_a^* \Delta_r + w_{13}^* B_{ai}^2 \Delta_a \Delta_r^*    \nonumber \\
&+ w_{14} B_{a0} B_{r0} \Delta_a^* \Delta_a + w_{15} B_{a0}^2 \Delta_a^* \Delta_r + w_{15}^* B_{a0}^2 \Delta_a \Delta_r^* + i w_{16} (\Delta_a^* \Delta_a)^2   \nonumber \\
& + {\rm i} w_{17} B_{ai}^2 \Delta_a^* \Delta_a + {\rm i} w_{18} B_{a0}^2 \Delta_a^* \Delta_a. \label{L_int}
\end{align}
Notice that, by $Z_2$ reflection symmetry, the coefficients $c_0$, $c_5$, $w_1$, $w_5$, $w_6$, $w_7$, $w_9$, $w_{10}$, $w_{14}$, $w_{16}$, $w_{17}$ and $w_{18}$ are purely real.

Then, the basic condition \eqref{Z_well-defined} requires
\begin{align}
w_5 \geq 0, \qquad w_{16} \geq 0.
\end{align}
Due to absence of time derivative terms in \eqref{L_int}, one may intuitively think the dynamical KMS symmetry would not adequately constrain the Lagrangian \eqref{L_int}. However, imposing the dynamical KMS symmetry, we still find interesting constraints
\begin{align}
& c_0 = c_1 = c_1^*, \quad \quad c_2 = c_2^*, \quad \quad c_3 = c_3^*, \quad \quad c_5 =2c_4 = 2c_4^*, \quad \quad w_0 = w_0^*,\nonumber \\
& w_2 = w_2^*, \quad \quad \quad w_3 = w_3^*, \quad \quad  w_4 =w_4^*, \quad \quad w_8 = w_8^*, \quad \quad w_{11} = w_{11}^*, \nonumber \\
& w_{12} = w_{12}^*, \quad \quad w_{13} = w_{13}^*,\quad \quad w_{15} = w_{15}^*.
\end{align}
Now putting together the constraints from $Z_2$ reflection symmetry and dynamical KMS symmetry, we find that all the coefficients (i.e., $c$'s and $w$'s) in \eqref{L_int} are purely real.

Finally, we briefly discuss constraint from Onsager relations, which are automatically satisfied at lower orders once dynamical KMS symmetry is imposed. However, there is one exception at quartic order. Interestingly, we found that the Onsager relations among $rrra$-terms \cite{Crossley:2015evo} give an additional constraint
\begin{align}
b_3 = -c_2, \label{onsager_b3c2}
\end{align}
which is useful in casting the EFT into stochastic equations \cite{Donos:2023ibv}. Interestingly, the result \eqref{onsager_b3c2} can actually be derived from local KMS condition \cite{Crossley:2015evo} imposed for generating functional $W = - {\rm i} \log Z$. Here, we shall imagine $\Delta$ as source for some operator $\psi_b$ as in \cite{Bu:2021clf} (which is more naturally motivated from holographic perspective). In addition, our holographic study will convince the constraint \eqref{onsager_b3c2}.

\subsection{Comment on the EFT}

Here, we make a brief comment on the EFT presented in last subsection.

Firstly, we check the dynamical modes described by the EFT action, which will be achieved by considering dynamical equations for physical fields $\varphi_r$, $\Delta_r$ and $\Delta_r^*$. Variation of the action $S_{eff}$ with respect to $a$-fields gives the dynamical equations
\begin{align}
\frac{\delta S_{eff}}{\delta \varphi_a} =0, \qquad  \frac{\delta S_{eff}}{\delta \Delta_a} =0, \qquad \frac{\delta S_{eff}}{\delta \Delta_a^*} =0. \label{eom_r-variable}
\end{align}

We proceed by considering the high temperature phase so that $\Delta_r$ does not have a background. In \eqref{eom_r-variable}, setting all $a$-fields to zero and ignoring nonlinear terms, in the hydrodynamic limit we obtain dispersion relations for dynamical modes
\begin{align}
\omega_{diff} = - {\rm i} D q^2 + \cdots, \qquad \omega_{\Delta} = -{\rm i} \Gamma_{\Delta} - {\rm i} D_{\Delta} q^2 + \cdots,
\end{align}
where
\begin{align}
D = - \frac{a_2}{a_0}, \qquad \Gamma_\Delta = \frac{b_0}{{\rm Re}(b_1)}, \qquad D_{\Delta} = \frac{b_3}{{\rm Re}(b_1)}. \label{D_Gamma}
\end{align}
Here, $\omega_{diff}$ is the diffusive mode for the conserved U(1) charge density. Notice that the parameter $b_0 \sim T_c - T$ near the critical point, and becomes negative in the symmetric phase. Thus, $\omega_{\Delta}$ represents the quasi-hydro mode associated with the order parameter (indeed its amplitude) when the system is slightly above the critical temperature.

The EFT can also be used to study dynamical modes when $T\lesssim T_c$ \cite{Donos:2023ibv}. In this case, the order parameter $\Delta_r$ will gain a background (i.e., the condensate). We make the following replacement in the action $S_{eff}$
\begin{align}
\Delta_r(x) \to \Delta_0 + \Delta(x), \qquad \Delta_r^*(x) \to \Delta_0 + \Delta^*(x)
\end{align}
where the condensate $\Delta_0$ is assumed to be a constant. Then, linearizing the dynamical equations \eqref{eom_r-variable}, one can obtain dispersion relations for dynamical modes
\begin{align}
\omega_{\rm H} = - {\rm i} \Gamma_{\rm H} - {\rm i} D_{\rm H} q^2 + \cdots, \qquad \omega_{\pm} = \pm c_s q - {\rm i} D_s q^2+ \cdots,
\end{align}
where various coefficients could be found in \cite{Donos:2023ibv}. Interestingly, a sound mode emerges due to spontaneously breaking of the global U(1) symmetry.

Secondly, we would like to clarify the relationship between the EFT above and relevant studies in the literature. Actually, the Model F of \cite{RevModPhys.49.435} corresponds to further truncating the EFT Lagrangian $\mathcal L_{eff}$ to the following one \cite{Donos:2023ibv}
\begin{align}
\mathcal L_{\rm MF} & = a_0 B_{a0} B_{r0} + a_2 B_{ai} \partial_0 B_{ri} + b_0 \Delta_a^* \Delta_r + b_0 \Delta_a \Delta_r^* + b_1 \Delta_a^* \partial_0 \Delta_r \nonumber \\
& + b_1^* \Delta_a \partial_0 \Delta_r^* + b_3 (\mathcal D_i \Delta_a)^* (\mathcal D_i \Delta_r) + b_3 (\mathcal D_i \Delta_a) (\mathcal D_i \Delta_r)^* + c_0 B_{a0} \Delta_r^* \Delta_r \nonumber \\
& + c_0 B_{r0} \Delta_a^* \Delta_r + c_0 B_{r0} \Delta_a \Delta_r^* + {\rm i} c_2 B_{ai} (\mathcal D_i \Delta_r)^* \Delta_r - {\rm i} c_2 B_{ai} (\mathcal D_i \Delta_r) \Delta_r^* \nonumber \\
&  + c_3 \Delta_a \Delta_r^* \Delta_r^* \Delta_r + c_3 \Delta_a^* \Delta_r \Delta_r^* \Delta_r - {\rm i} \beta^{-1} a_2 B_{ai}^2 -2 {\rm i} \beta^{-1} {\rm Re}(b_1) \Delta_a^* \Delta_a
\end{align}
So, comparing $\mathcal L_{\rm MF}$ to $\mathcal L_{eff}$, the EFT we wrote down stands for a significant extension of relevant results in the literature \cite{RevModPhys.49.435,Donos:2023ibv}, particularly on the treatment of thermal fluctuations. On the one hand, on top of Gaussian white noises (denoted by $u_0$-, $u_1$- and $v_0$-terms), we have added higher derivative corrections, such as $u_{3-8}$-terms and $v_{1-3}$-terms. The latter can be understood as Gaussian but coloured noises. On the other hand, regarding the interaction part $\mathcal L_{int}$, the $w_{0-9}$-terms can be thought of as multiplicative noises, while the $w_{10-18}$-terms represent non-Gaussian noises. With the technique of \cite{Lin:2023bli}, this will become more transparent by converting $w_{10-18}$-terms into stochastic forces obeying non-Gaussian distributions.

Through dynamical KMS symmetry, all these corrections (i.e., Gaussian coloured noises, multiplicative noises or non-Gaussian noises) shall be accompanied by suitable higher time-derivative terms that we decided not to pursue in present work. For the example of charge diffusions, this has been intensively investigated in \cite{Crossley:2015evo,Jain:2020zhu,Bu:2022esd}.

\section{Holographic derivation of EFT action} \label{EFT_holo}

In this section we provide a holographic derivation of the EFT action presented in section \ref{EFT_construction}. To this end, we consider the minimal holographic superfluid model \cite{Hartnoll:2008vx,Hartnoll:2008kx}, which consists of a scalar QED in an asymptotically AdS$_5$ black brane. The total action is
\begin{align}
    S = S_0 + S_{\rm bdy}
\end{align}
where the bulk action $S_0$ is
\begin{equation}
    S_0 = \int d^{5}x\sqrt{-g}\left[-\frac{1}{4}F^{MN}F_{MN}-
    D_M \Psi \left(D^M \Psi \right)^* -m_0^2 \Psi^* \Psi \right] \label{S0}
\end{equation}
where $D_M = \nabla_M - i A_M$. We use a $^*$ to denote charge conjugate. The term $S_{\rm bdy}$ depends on specific boundary conditions for bulk fields and will be specified later. We will take $m_0^2 =- 4$ so that analytical solutions for bulk fields become possible \cite{Herzog:2010vz}. The bulk theory is invariant under the U(1) gauge transformation:
\begin{align}
    A_M \to A_M + \nabla_M \Lambda(r,x^\mu), \qquad \Psi \to \Psi e^{i \Lambda(r,x^\mu)}, \qquad \Psi^* \to \Psi^* e^{-i \Lambda(r,x^\mu)} \label{gauge_symmetry}
\end{align}
which will play a crucial role in subsequent analysis.

\begin{figure} [htbp!]
		\centering
		\begin{tikzpicture}[]
		
		\draw[cyan, ultra thick] (-5.5,0)--(-2.8,0);
		\draw[cyan, ultra thick] (-1.2,0)--(1.5,0);
		\draw[cyan, ultra thick] (-2.81,0.019) arc (-180:0:0.8);
		\draw[cyan, ->,very thick] (-1,0)--(0.2,0);
		\draw[cyan, ->,very thick] (-4.2,0)--(-4,0);
		\draw[fill] (-5.5,0) circle [radius=0.05];
		\node[above] at (-5.5,0) {\small $\infty_2$};
		\draw[fill] (1.5,0) circle [radius=0.05];
		\node[above] at (1.5,0) {\small $\infty_1$};
		\draw[fill,red] (-2,0) circle [radius=0.05];
		\node[above] at (-2,0) {\small $r_h$};				
		
		\draw[-to,thick] (1.85,0)--(2.45,0);

		\node[below] at (9.5,-0.1) {\small Re$(r)$};
		\draw[->,ultra thick] (2.9,0)--(9.6,0);
		\node[right] at (3.8,1) {\small Im$(r)$};
		\draw[->,ultra thick] (3.6,-1)--(3.6,1.2);
		\draw[cyan, very thick] (5.2,-0.18) arc (165:-165:-0.7);
		\draw[cyan, very thick] (5.2,0.2)--(8.3,0.2);
		\draw[cyan, very thick] (5.2,-0.2)--(8.3,-0.2);
		\draw[cyan, ->,very thick] (6,-0.2)--(7,-0.2);
		\draw[cyan, <-,very thick] (6,0.2)--(7,0.2);
		\draw[fill] (3.6,0) circle [radius=0.05];
		\draw[fill] (8.3,0.2) circle [radius=0.05];
		\node[below] at (8.3,-0.15) {\small $\infty_1 $};
		\draw[fill] (8.3,-0.2) circle [radius=0.05];
		\node[above] at (8.3,0.15) {\small $\infty_2 $};
		\draw[fill , red ] (4.5,0) circle [radius=0.07];
		\node[below] at (4.5,0) {\footnotesize $r=r_h $};
		\draw[cyan, thick,<->] (4.52,0.08)--(4.8,0.62);
		\node[above] at (4.5, 0.2) {\footnotesize $\epsilon$};
		\end{tikzpicture}
		\caption{From complexified (analytically continued near horizon) double AdS (left) \cite{Crossley:2015tka} to the holographic SK contour (right) \cite{Glorioso:2018mmw}. Indeed, the two horizontal legs overlap with the real axis.} \label{rcontour}
\end{figure}
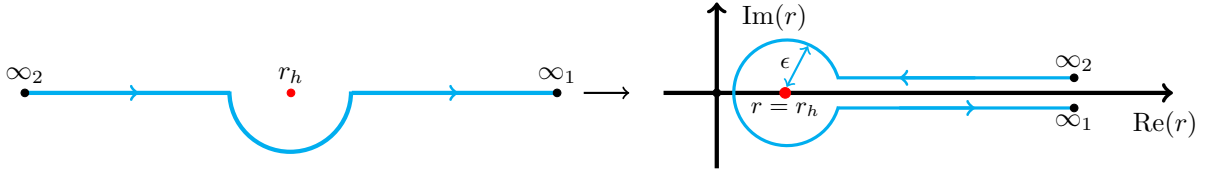

We will work in the probe limit. Then, in the ingoing Eddington-Finkelstein (EF) coordinate system, the metric of AdS$_5$ black brane is given by
\begin{align}
    ds^2 = g_{MN} dx^M dx^N = 2 dv dr - f(r) r^2 dv^2 + r^2 \delta_{ij} dx^i dx^j
\end{align}
where $f(r) = 1- r_h^4/r^4$. Here, $r=r_h$ is the location of event horizon and $r= \infty $ is the AdS boundary. Practically, we will take $r_h =1$ for convenience. Following the prescription of \cite{Glorioso:2018mmw}, a holographic dual for Schwinger-Keldysh closed time contour is obtained by analytically continuing the radial coordinate $r$ around the horizon and then doubling it, see Figure \ref{rcontour}.

\subsection{Holographic program towards boundary EFT} \label{holo_dictionary}

In this section, we explain the strategy of deriving boundary effective action from the dynamics of bulk theory, which amounts to a holographic RG program. Such a program was initiated in \cite{Crossley:2015tka} for a pure AdS gravity (see also \cite{Nickel:2010pr}), and later revisited in \cite{Bu:2020jfo,Bu:2021clf,Baggioli:2023tlc}.

The starting point is the holographic dictionary \cite{Gubser:1998bc,Witten:1998qj}
\begin{align}
    Z_{\rm AdS} = Z_{\rm CFT}.
\end{align}
The partition function $Z_{\rm CFT}$ is expressed as a path integral over the low energy modes (collectively denoted by $X$) for the boundary theory,
\begin{align}
    Z_{\rm CFT} = \int [D X] e^{i S_{eff}[X]}, \label{Z_CFT}
\end{align}
which, once identified with \eqref{Z_EFT}, tells that $S_{eff}$ is the boundary effective action to be derived through bulk calculations. On the other hand, the bulk partition function $Z_{\rm AdS}$ is
\begin{align}
    Z_{\rm AdS} & = \int [D A_M^\prime] [D \Psi^\prime] [D\Psi^{\prime*}]  e^{i S_0[A_M^\prime, \, \Psi^\prime, \, \Psi^{\prime*}] + i S_{\rm bdy}} \nonumber \\
    & = \int [D\Lambda] [D A_\mu] [D \Psi] [D\Psi^*]  e^{i S_0[A_\mu, \, \Psi, \, \Psi^*] + i S_{\rm bdy}} \label{Z_AdS}
\end{align}
where the primed field configuration $(A_M', \Psi', \Psi^{\prime *})$ means no gauge-fixing, while $(A_\mu, \Psi, \Psi^*)$ denotes bulk field configuration with a specific gauge choice. The missed gauge degree of freedom arising from gauge-fixing $(A_M', \Psi', \Psi^{\prime *})$ to be $(A_\mu, \Psi, \Psi^*)$ will be captured by the gauge transformation parameter $\Lambda$. In other words, the radial component $A_r$ is fixed via a certain gauge choice, i.e., $A_r = A_r[A_\mu]$, and its dynamics will be equivalently described by the gauge transformation parameter $\Lambda$.

Now, we would like to cast \eqref{Z_AdS} into the desired form \eqref{Z_CFT}. This amounts to identifying holographic duals of the low energy modes for boundary theory and integrating out those heavy modes in the bulk. To this end, we consider near-boundary behavior of bulk fields
\begin{align}
    &A_\mu'(r \to \infty, x^\alpha) \to  \mathcal A_\mu(x^\alpha) + \frac{\partial_v \mathcal{A}_{\mu}}{r} - \frac{1}{2} \partial^\nu \mathcal F_{\mu\nu} \frac{\log r}{r^2} + \frac{J_\mu'(x^\alpha)}{r^2} + \cdots, \nonumber \\
    &\Psi'(r \to \infty,x^\alpha) \to \psi_b'(x^\alpha) \frac{\log r}{r^2} + \frac{O(x^\alpha)}{r^2} + \cdots, \nonumber \\
    &\Psi^{\prime *}(r \to \infty,x^\alpha) \to \psi_b^{\prime *}(x^\alpha) \frac{\log r}{r^2} + \frac{O^*(x^\alpha)}{r^2} + \cdots. \label{near_bounday_gauge_fixed}
\end{align}
Recall that $\mathcal A_\mu$ is an external gauge potential for the boundary theory. Through the gauge transformation \eqref{gauge_symmetry}, we easily obtain
\begin{align}
    &A_\mu(r \to \infty, x^\alpha) \to  B_\mu(x^\alpha) + \frac{\partial_v B_{\mu}}{r} - \frac{1}{2} \partial^\nu \mathcal F_{\mu\nu} \frac{\log r}{r^2} + \frac{J_\mu(x^\alpha)}{r^2}  + \cdots, \nonumber \\
    &\Psi(r \to \infty,x^\alpha) \to \psi_b(x^\alpha) \frac{\log r}{r^2} + \frac{\Delta(x^\alpha)}{r^2} + \cdots, \nonumber \\
    &\Psi^*(r \to \infty,x^\alpha) \to \psi_b^*(x^\alpha) \frac{\log r}{r^2} + \frac{\Delta^*(x^\alpha)}{r^2} + \cdots,
\end{align}
where $B_\mu \equiv \mathcal A_\mu + \partial_\mu \varphi$, $\psi_b = \psi_b' e^{i\varphi}$, and $\Delta = O e^{i \varphi}$. Here, $\varphi$ is the boundary value of the bulk gauge transformation parameter, $\varphi \equiv \Lambda(r=\infty)$. According to \cite{Crossley:2015evo,Nickel:2010pr,Glorioso:2018mmw}, we interpret $\varphi$ as the diffusive field associated with the conserved $U(1)$ charge on the boundary. While the physics of order parameter can be described by either $O$ or $\Delta$, we find it more natural to work with $\Delta$ since the holographic calculations will be carried out in a specific gauge choice.

Near the critical point, both the charge density described by $\varphi$ and the order parameter $\Delta$ shall be retained in the low energy EFT. Therefore, once the bulk components $A_\mu, \Psi, \Psi^*$ (dual to heavy modes of boundary theory) are integrated out, \eqref{Z_AdS} is cast into the following desired form
\begin{align}
    Z_{\rm AdS} = \int [D \varphi] [ D \Delta] [D \Delta^*] e^{i S_0|_{\rm p.o.s} + i S_{\rm bdy}}
\end{align}
Here, we have utilized saddle point limit of holographic dictionary so that $S_0|_{\rm p.o.s}$ stands for the partially on-shell bulk action by substituting bulk solution for $A_\mu, \Psi, \Psi^*$ in the bulk action:
\begin{align}
    S_0|_{\rm p.o.s} = S_0 \left[ A_\mu[B_\mu, \Delta, \Delta^*], \Psi[B_\mu, \Delta, \Delta^*], \Psi^*[B_\mu, \Delta, \Delta^*] \right]
\end{align}
Particularly, by partially on-shell, when solving the bulk fields, we will not impose the constraint equation so that $\varphi$ is kept dynamical and un-integrated out. For the scalar field $\Psi$, the boundary condition will be to fix the normalizable mode so that $\Delta$ becomes a dynamical field on the boundary. Throughout this work, we choose the following gauge choice
\begin{align}
     A_r = - \frac{A_v}{r^2f(r)}  \label{radial_gauge_EF}
\end{align}


In the saddle point approximation, the derivation of boundary effective action reduces to solving bulk dynamics in the partially on-shell sense. This prescription will become more natural if we revisit the bulk variational problem based on the gauge-fixed configuration $(A_\mu, A_r[A_\mu], \Psi, \Psi^*)$, which we will explain below.

Since the field configuration $(A_M', \Psi', \Psi^{*'})$ does not assume any gauge-fixing, they can be varied freely,
\begin{align}
    A_M' \to A_M' + \delta A_M', \qquad \Psi' \to \Psi' + \delta \Psi', \qquad \Psi^{\prime *} \to \Psi^{\prime *} + \delta \Psi^{\prime *}
\end{align}
Then, the variation of bulk action is
\begin{align}
    \delta S_0 = \int d^5x \sqrt{-g} &\left\{ \left( \nabla_M F^{\prime MN} - \mathcal J^{\prime N} \right) \delta A_N' + \left(D_M' D^{\prime M} \Psi' - m_0^2 \Psi' \right)^* \delta \Psi' \right. \nonumber \\
    & \quad \left. + \left(D_M' D^{\prime M} \Psi' - m_0^2 \Psi' \right) \delta \Psi^{\prime *}   \right\} + S_\partial, \label{deltaS0}
\end{align}
where $D_M' = \nabla_M - i A_M'$, and $S_\partial$ is a potential boundary term which will not be crucial in subsequent analysis. The bulk current $\mathcal J_M'$ is
\begin{align}
    \mathcal J_M' = i \left[ \Psi^{\prime *} (\nabla_M - i A_M') \Psi' - \Psi' (\nabla_M + i A_M') \Psi^{\prime *} \right]
\end{align}

Actually, a field configuration with specific gauge-fixing is achieved through a gauge transformation
\begin{align}
    A_r = A_r' + \nabla_r \Lambda, \qquad A_\mu = A_\mu' + \nabla_\mu \Lambda, \qquad \Psi = e^{i \Lambda} \Psi', \qquad \Psi^* = e^{- i \Lambda} \Psi^{\prime *},
\end{align}
which, together with the radial gauge choice \eqref{radial_gauge_EF}, tells
\begin{align}
    &\delta A_r' = -\frac{\delta A_v}{r^2f(r)} - \nabla_r \delta \Lambda, \qquad \qquad \, \,  \delta A_\mu' = \delta A_\mu - \nabla_\mu \delta \Lambda, \nonumber \\ &\delta \Psi' = e^{-i \Lambda} \delta \Psi -i e^{-i \Lambda} \Psi \delta \Lambda, \qquad \quad \delta \Psi^{\prime *} = e^{i \Lambda} \delta \Psi^* + i e^{i \Lambda} \Psi^* \delta \Lambda. \label{varition_gauge1-2}
\end{align}
Due to the gauge-fixing \eqref{radial_gauge_EF}, we cannot freely vary $A_r$ any longer, i.e., $\delta A_r = - \delta A_v/(r^2f(r))$. However, we can freely have $\delta \Lambda \neq 0$. Then, with the help of \eqref{varition_gauge1-2}, we could express $\delta S_0$ in \eqref{deltaS0} in terms of gauge-fixed configuration. Eventually, from $\delta S_0$, we obtain the dynamical components of bulk equations of motion (EOMs):
\begin{align}
    &\delta A_v \neq 0 \Rightarrow \nabla_M F^{Mv} - g^{v A} \mathcal J_A - \frac{1}{r^2f(r)} \left( \nabla_M F^{Mr} - g^{r A} \mathcal J_A \right) =0, \nonumber \\
    &\delta A_i \neq 0 \Rightarrow \nabla_M F^{Mi} - g^{iA}\mathcal J_A = 0, \nonumber \\
    &\delta \Psi^* \neq 0 \Rightarrow D_M D^M \Psi - m_0^2 \Psi =0 , \nonumber \\
    &\delta \Psi \neq 0 \Rightarrow (D_M D^M \Psi)^* - m_0^2 \Psi^* =0, \label{dynamical_EOMs}
\end{align}
and the contracted Bianchi identity
\begin{align}
    \delta \Lambda \neq 0 \Rightarrow \nabla_N (\nabla_M F^{MN} - \mathcal J^N) = 0. \label{Bianchi_ID}
\end{align}
Lastly, we would have a boundary term,
\begin{align}
    S_\partial = \int d^4x \sqrt{-\gamma} n_N \left[ -\nabla_M F^{MN} + \mathcal J^M \right] \delta \Lambda|_{\rm bdy}
\end{align}
which would give the constraint equation if the gauge transformation parameter could be varied on the boundary
\begin{align}
    \delta \Lambda|_{\rm bdy} \neq 0 \Rightarrow \nabla_M F^{Mr} - \mathcal J^r \big|_{\rm bdy} =0 \label{constraint_bdy}
\end{align}
The bulk current $\mathcal J^M$ is
\begin{align}
\mathcal J_M = i \left[ \Psi^* (\nabla_M - i A_M) \Psi - \Psi (\nabla_M + i A_M) \Psi^* \right]
\end{align}

Obviously, under radial gauge choice \eqref{radial_gauge_EF}, the dynamical EOMs \eqref{dynamical_EOMs} fully solve the bulk fields. Then, the quantity $\nabla_M F^{Mr} - \mathcal J^r$ entering the constraint equation is known at any spacetime point. Notice that, due to the Bianchi identity \eqref{Bianchi_ID}, the constraint $\nabla_M F^{Mr} - \mathcal J^r$ will vanish at any $r$-slice once the dynamical EOMs \eqref{dynamical_EOMs} are satisfied.

%

\subsection{Boundary conditions and the boundary term $S_{\rm bdy}$}

Recall that, as explained in subsection \ref{holo_dictionary}, the boundary data will be $B_\mu$ and $\Delta$, which are actually the dynamical fields for the boundary theory. Thus, at the AdS boundary, we will impose Dirichlet conditions for $A_\mu$ so that its boundary value will be fixed to $B_\mu$. In contrast, we will impose a Neumann-type boundary condition for $\Psi$ such that its normalizable mode will be fixed to $\Delta$.

Now it is ready to specify the boundary term $S_{\rm bdy}$, which will play two roles: remove UV divergences in the bulk action $S_0$ as $r \to \infty$; guarantee the bulk variational problem to be well-posed. Without presenting the derivation, we just take the boundary term from \cite{Bu:2021clf} and write it here for later convenience
\begin{align}
S_{\rm bdy}= \int d^4x \sqrt{-\gamma} \left\{- \frac{1}{4} F_{\mu\nu} F^{\mu\nu} \log r + 2 \Psi^* \Psi - \frac{\Psi^* \Psi}{\log r} + n_M \left( \Psi^* \nabla^M \Psi + \Psi \nabla^M \Psi^* \right) \right\}
\end{align}
where $\gamma$ is determinant of the induced metric on a constant $r$-surface with $r\to \infty$ taken eventually. Then, it is straightforward to check that the variation of total bulk action takes an expected form
\begin{align}
    \delta(S_0 + S_{\rm bdy}) = \int d^4x \left[ (J^\mu + \cdots)\delta B_\mu + \psi_b\delta \Delta^* + \psi_b^* \delta \Delta \right]
\end{align}
where $\cdots$ are possible contact terms.

However, in order to fully determine time-component of bulk gauge field $A_v$, we need an additional boundary condition. Physically, such a condition corresponds to chemical shift symmetry \eqref{chemical_shift}-\eqref{chemical_shift1} for the boundary theory:
\begin{align}
    \varphi_r \to \varphi_r + \sigma(\vec x), \qquad \varphi_a \to \varphi_a, \qquad \Delta_r \to e^{{\rm i} \sigma(\vec x)} \Delta_r, \qquad \Delta_a \to e^{{\rm i} \sigma(\vec x)} \Delta_a. \label{chmeical_shift2}
\end{align}
This claim will become transparent if we re-consider the bulk gauge symmetry \eqref{gauge_symmetry}. Notice that, after the radial gauge-fixing \eqref{radial_gauge_EF}, we still have a residual gauge symmetry
\begin{align}
    A_\mu \to A_\mu + \partial_\mu \Lambda(x^\alpha), \qquad \Psi \to e^{i \Lambda(x^\alpha)}\Psi, \qquad \Psi^* \to e^{-i \Lambda(x^\alpha)} \Psi^*, \label{residual_gauge_symmetry}
\end{align}
where the gauge parameter $\Lambda$ now depends on boundary coordinates only. If we further gauge-fix $A_v$ by, for instance,  taking \cite{Glorioso:2018mmw}
\begin{align}
    A_v(r=r_h-\epsilon) =0 \label{horizon_vanishing}
\end{align}
the residual gauge symmetry \eqref{residual_gauge_symmetry} breaks down to the case of $\Lambda = \Lambda(\vec x)$. Thus,
its boundary version will be exactly that of \eqref{chmeical_shift2}.

In fact, we could have put some generic $x^\alpha$-dependent function on the right-handed side of \eqref{horizon_vanishing}, i.e, taking a more generic gauge-fixing $A_v(r=r_h) = F(x^\alpha)$. This general treatment also does the job of breaking boundary version of \eqref{residual_gauge_symmetry} to the chemical shift symmetry \eqref{chmeical_shift2}. It is tempting to interpret that different choice of $F(x^\alpha)$ corresponds to different frame. We will leave such an exploration as a future task.

Before concluding this subsection, we simplify the bulk action $S_0$ a bit by utilizing the dynamical EOMs \eqref{dynamical_EOMs}. After integrating by parts in \eqref{S0}, we obtain
\begin{align}
    S_0 & = \int d^4x \sqrt{-\gamma} n_r \left\{- \frac{1}{2} A_v F^{rv} - \frac{1}{2} A_i F^{ri} - \frac{1}{2} \Psi^* (g^{rr} \partial_r \Psi + g^{rv} \partial_v \Psi) \right. \nonumber \\
    & \qquad \qquad \qquad \qquad  \left. - \frac{1}{2} \Psi (g^{rr} \partial_r \Psi^* + g^{rv} \partial_v \Psi^*) \right\} \Bigg|_{r=\infty_2}^{r=\infty_1} \nonumber \\
    &+ \int d^5x \sqrt{-g}  \left( \frac{1}{2} A_N \nabla_M F^{MN} \right).
\end{align}
With the near-boundary behavior for bulk fields \eqref{near_bounday_gauge_fixed}, we eventually obtain
\begin{align}
    S_{eff} =& \int d^4x \Big[ \frac{1}{2} B_v \partial_v^2 B_v - \frac{1}{2} B_i \partial_v^2 B_i - \frac{1}{4} B_v \partial^{\nu} \mathcal{F}_{v \nu} + \frac{1}{4} B_i \partial^{\nu} \mathcal{F}_{i \nu} \nonumber \\
    & \qquad \qquad - B_v J_v + B_i J_i - \frac{1}{2} \psi_b \Delta^* - \frac{1}{2} \psi_b^* \Delta \Big] \Bigg|_2^1 \nonumber  \\
    &+ \int d^5x \sqrt{-g} \left\{ -\frac{i}{2 r^2 f(r)} A_v [\Psi^* \partial_v \Psi - \Psi \partial_v \Psi^*] + \frac{i}{2 r^2} A_i [\Psi^* \partial_i \Psi - \Psi \partial_i \Psi^*]  \right. \nonumber \\
    & \qquad \qquad \qquad \quad \left. - \frac{1}{r^2 f(r)}A_v^2 \Psi \Psi^* + \frac{1}{r^2} A_i^2 \Psi \Psi^* \right\}, \label{Seff_bulk}
\end{align}
where we made use of the radial gauge choice \eqref{radial_gauge_EF}.
%

\subsection{Holographic calculation}

In this subsection we set up a perturbative approach and solve the dynamical EOMs \eqref{dynamical_EOMs} on the radial contour of Figure \ref{rcontour}. Plugging the perturbative solutions into the bulk action \eqref{Seff_bulk}, we obtain the EFT action of section \ref{EFT_construction} as well as holographic results for various coefficients.

First, we create a finite density state in the high temperature phase, which corresponds to the following static background for bulk fields
\begin{align}
    \bar A_v = \phi(r), \qquad  \bar A_r = - \frac{\phi(r)}{r^2 f(r)}, \qquad \bar \Psi = \bar \Psi^* = 0,  \label{static_ansatz}
\end{align}
where $\phi$ is known analytically
\begin{align}
    \phi = \mu \left(1-\frac{1}{r^2}\right)
\end{align}
Here, $\mu$ has the physical meaning of chemical potential. It was realized that \cite{Herzog:2010vz} only when $\mu =2$, can one obtain analytical solutions for bulk perturbation to be introduced later. Thus, throughout this work, we will take
\begin{align}
    \mu = \mu_0 + \delta \mu, \qquad {\rm with} \qquad \mu_0 =2,
\end{align}
where $\delta \mu$ represents a perturbation to the critical chemical potential $\mu_0$, which drives the system a little bit away from the critical point.

Then, on top of the background \eqref{static_ansatz}, we turn on general perturbations so that the bulk fields are
\begin{align}
&A_r =- \frac{\phi(r) + \alpha_v(r, x^\alpha)}{r^2 f(r)},\qquad \qquad A_\mu = \phi(r) \delta_{v \mu} + \alpha_\mu(r, x^\alpha), \nonumber \\
&\Psi = 0 + \psi(r, x^\alpha), \qquad \qquad \qquad \quad \Psi^* = 0 + \psi^*(r,x^\alpha),
\end{align}
where $A_r$ is completely fixed by the gauge convention \eqref{radial_gauge_EF}. In terms of bulk perturbation, the dynamical EOMs \eqref{dynamical_EOMs} read
\begin{align}
    0= & \partial_r ( r^3 \partial_r \alpha_v) + \frac{2r}{f(r)} \partial_r \partial_v \alpha_v + \left[ \frac{1}{f(r)} - \frac{ r \partial_r f(r)}{ f^2(r)} \right] \partial_v \alpha_v + \frac{1}{r f^2(r)} \partial_v^2 \alpha_v \nonumber \\
    & + \frac{1}{r f(r)} ({\vec \partial}^{\,2} \alpha_v - \partial_v \partial_i \alpha_i ) - \frac{{\rm i}r}{ f(r)} \left[ \psi^* \partial_v \psi - \psi \partial_v \psi^* -2{\rm i} \left(\phi +\alpha_v \right) \psi \psi^* \right], \label{alphav_eom} \\
    0= & \partial_r\left[r^3 f(r) \partial_r \alpha_i \right]+ 2r \partial_r \partial_v \alpha_i + \partial_v \alpha_i  + \frac{1}{rf(r)} \partial_v \partial_i \alpha_v + \frac{1}{r} (\partial_k ^{2} \alpha_i - \partial_i \partial_k \alpha_k) \nonumber \\
    & - {\rm i} r \left[ \psi^* \partial_i \psi - \psi \partial_i \psi^* - 2 {\rm i} \alpha_i \psi \psi^* \right], \label{alphai_eom} \\
    0= & \partial_r \left[r^{5} f(r) \partial_r \psi \right] + 2r^3 \partial_r \partial_v \psi + 3 r^{2} \partial_v \psi + r {\vec\partial}^{\,2} \psi + \frac{2{\rm i}r}{f(r)} (\phi + \alpha_v) \partial_v \psi \nonumber \\
    & + \frac{{\rm i}r}{f(r)} \psi \partial_v \alpha_v - 2{\rm i} r \alpha_i \partial_i \psi - {\rm i}r \psi \partial_i \alpha_i + \frac{r \phi^2}{f(r)} \psi + \frac{r}{f(r)} (2\phi \alpha_v + \alpha_v^2) \psi\nonumber \\
    & -r \alpha_i^2 \psi - m_0^2 r^3 \psi, \label{psi_eom} \\
    0= & \partial_r \left[r^{5} f(r) \partial_r \psi^* \right] + 2r^3 \partial_r \partial_v \psi^* + 3 r^{2} \partial_v \psi^* + r {\vec\partial}^{\,2} \psi^* - \frac{2{\rm i}r}{f(r)} (\phi + \alpha_v) \partial_v \psi^* \nonumber \\
    & - \frac{{\rm i}r}{f(r)} \psi^* \partial_v \alpha_v + 2{\rm i} r \alpha_i \partial_i \psi^* + {\rm i}r \psi^* \partial_i \alpha_i + \frac{r \phi^2}{f(r)} \psi^* + \frac{r}{f(r)} (2\phi \alpha_v + \alpha_v^2) \psi^* \nonumber \\
    & -r \alpha_i^2 \psi^* - m_0^2 r^3 \psi^*,  \label{psi*_eom}
\end{align}
which form a system of nonlinear partial differential equation (PDEs). Generally, it is challenging to solve these equations. Nevertheless, in accord with the spirit of EFT, we will solve these PDEs by adopting several approximations. It turns out that we need a triple expansion.

First, we implement a derivative expansion:
\begin{align}
    \alpha_v & = \xi^0 \alpha_v^{(0)} + \xi^1 \alpha_v^{(1)} + \xi^2 \alpha_v^{(2)} + \cdots, \qquad \quad
    \alpha_i = \xi^0 \alpha_i^{(0)} + \xi^1 \alpha_i^{(1)} + \xi^2 \alpha_i^{(2)} + \cdots, \nonumber \\
    \psi & = \xi^0 \psi^{(0)} + \xi^1 \psi^{(1)} + \xi^2 \psi^{(2)} + \cdots, \qquad \quad
    \psi^*  = \xi^0 \psi^{*(0)} + \xi^1 \psi^{*(1)} + \xi^2 \psi^{*(2)} + \cdots, \label{xi-expansion}
\end{align}
where $\xi \sim \partial_\mu$. Physically, such an expansion corresponds to the assumption that the boundary system evolves slowly in the hydrodynamic limit. The derivative expansion renders the system of PDEs into a nonlinear system of ordinary differential equations (ODEs).

Second, we make an expansion in the number of boundary data $B_\mu$, $\Delta$ and $\Delta^*$:
\begin{align}
    \alpha_v^{(l)} & = \lambda^1 \alpha_v^{(l)(1)} + \lambda^2 \alpha_v^{(l)(2)} + \cdots, \qquad \quad
    \alpha_i^{(l)}  = \lambda^1 \alpha_i^{(l)(1)} + \lambda^2 \alpha_i^{(l)(2)} + \cdots, \nonumber \\
    \psi^{(l)} & = \lambda^1 \psi^{(l)(1)} + \lambda^2 \psi^{(l)(2)} + \cdots, \qquad \quad
    \psi^{*(l)} = \lambda^1 \psi^{*(l)(1)} + \lambda^2 \psi^{*(l)(2)} + \cdots,
\end{align}
where the expansion parameter $\lambda$ scales as $\lambda \sim B_\mu \sim \Delta \sim \Delta^*$. Importantly, through such an amplitude expansion, the nonlinear system of ODEs obtained via the derivative expansion is reduced to a decoupled linear system of ODEs.

Last, we will carry out an expansion in terms of chemical potential perturbation $\delta \mu$:
\begin{align}
    \alpha_v^{(l)(m)} & =  \alpha_v^{(l)(m)(0)} + \kappa \, \alpha_v^{(l)(m)(1)} + \cdots, \qquad
   \alpha_i^{(l)(m)}  =  \alpha_i^{(l)(m)(0)} + \kappa \, \alpha_i^{(l)(m)(1)} + \cdots, \nonumber \\
    \psi^{(l)(m)} & =  \psi^{(l)(m)(0)} + \kappa \, \psi^{(l)(m)(1)} + \cdots, \qquad
    \psi^{*(l)(m)} =  \psi^{*(l)(m)(0)} + \kappa \, \psi^{*(l)(m)(1)} + \cdots,
\end{align}
where $\kappa \sim \delta \mu$. Via this last expansion, analytical solutions for bulk perturbations become possible. Recall that $b_0$ in \eqref{L_Delta} will vanish at the critical point (see comment below \eqref{D_Gamma}). Thus, for the $\delta\mu$-correction, we will merely consider $\alpha_\mu^{(0)(1)(1)}$, $\psi^{(0)(1)(1)}$ and $\psi^{*(0)(1)(1)}$, which are relevant for the computation of $b_0$-term in \eqref{L_Delta}. In terms of EFT Lagrangian \eqref{L_diff}, \eqref{L_Delta} and \eqref{L_int}, we will track leading $\delta\mu$-correction to coefficients $b_0, v_0$, while compute all the rest coefficients exactly at the critical point.

Thanks to the triple expansion, the original nonlinear PDEs \eqref{alphav_eom}-\eqref{psi*_eom} are reduced into a decoupled set of ODEs
\begin{align}
    &\Box_v \alpha_v^{(l)(m)(n)}  = j_v^{(l)(m)(n)}, \qquad \qquad  \Box_i \alpha_i^{(l)(m)(n)}  = j_i^{(l)(m)(n)}, \nonumber \\
    &\Box_\psi \psi^{(l)(m)(n)} = j_\psi^{(l)(m)(n)},  \qquad \qquad \Box_\psi \psi^{*(l)(m)(n)} = j_{\psi^*}^{(l)(m)(n)},  \label{eom_triple_expansion}
\end{align}
where various differential operators are
\begin{align}
    \Box_v \equiv \partial_r \left( r^{3} \partial_r \right), \qquad \Box_i \equiv  \partial_r \left[ r^{3} f(r) \partial_r \right], \qquad
    \Box_\psi \equiv \partial_r \left[ r^{5} f(r) \partial_r \right] + \frac{ r \phi_0^2}{ f(r)} - m_0^2 r^3,
\end{align}
where $\phi_0 = \mu_0 (1 - 1/r^2)$. Explicit expressions for various source terms in \eqref{eom_triple_expansion} are collected in appendix \ref{source_terms}.

The AdS boundary conditions, cf. \eqref{near_bounday_gauge_fixed}, will be implemented in the following manner
\begin{align}
&\alpha_\mu^{(0)(1)(0)}(r=\infty_s) = B_{s\mu}, \qquad \qquad \qquad \quad \alpha_\mu^{(l)(m>1)(n)}(r=\infty_s) = 0, \nonumber \\
&\psi^{(0)(1)(0)}(r \to \infty_s) = \cdots + \frac{\Delta_s}{r^2} + \cdots, \qquad \psi^{(l)(m>1)(n)}(r \to \infty_s) = \cdots + \frac{0}{r^2} + \cdots, \nonumber \\
&\psi^{*(0)(1)(0)}(r \to \infty_s) = \cdots + \frac{\Delta_s^*}{r^2} + \cdots, \qquad \psi^{*(l)(m>1)(n)}(r \to \infty_s) = \cdots + \frac{0}{r^2} + \cdots, \label{AdS_condition_perturb}
\end{align}
while the horizon condition \eqref{horizon_vanishing} will be imposed at each order
\begin{align}
\alpha_v^{(l)(m)(n)}(r=r_h - \epsilon) = 0.
\end{align}

Here, we outline the strategy of solving \eqref{eom_triple_expansion}. For the time-component of bulk gauge field, we solve them by direct integration over the radial coordinate\footnote{At certain order, we would have to make redefinition over the source term so that the inner integral will behave well near the AdS boundary.}
\begin{align}
\alpha_{sv}^{(l)(m)(n)}(r) = \int_{\infty_s}^r \left[ \frac{1}{x^3} \int_{\infty_s}^x j_v^{(l)(m)(n)}(y) dy + \frac{c_s^{(l)(m)(n)}}{x^3} \right] dx + d_s^{(l)(m)(n)}, \label{alphav_lmn}
\end{align}
where the integration constants $c_s^{(l)(m)(n)}$ and $d_s^{(l)(m)(n)}$ could be determined by the boundary conditions summarized above.

For $\alpha_i$, $\psi$ and $\psi^*$, the lower order solutions could be written in compact forms. For instance, at the lowest order, we have
\begin{align}
&\alpha_i^{(0)(1)(0)}(r) = B_{2i} + B_{ai} \log \frac{r^2-r_h^2}{r^2 + r_h^2}, \nonumber \\
&\psi^{(0)(1)(0)} (r) = \frac{\Delta_2}{r^2 + r_h^2} - \frac{\Delta_a}{2 {\rm i} \pi} \frac{\log r - \log(r^2 - r_h^2)}{r^2 + r_h^2}, \nonumber \\
&\psi^{*(0)(1)(0)} (r) = \frac{\Delta_2^*}{r^2 + r_h^2} - \frac{\Delta_a^*}{2 {\rm i} \pi} \frac{\log r - \log(r^2 - r_h^2)}{r^2 + r_h^2}.
\end{align}
By virtue of Green's function method, the solutions for higher order perturbations could be written compactly,
\begin{align}
& \alpha_i^{(l)(m)(n)}(r) = \int_{\infty_2}^{\infty_1} G_\alpha(r,r') j_i^{(l)(m)(n)}(r') \, dr' , \nonumber \\
&\psi^{(l)(m)(n)}(r) = \int_{\infty_2}^{\infty_1} G_\psi(r,r') j_\psi^{(l)(m)(n)}(r') \, dr' , \nonumber \\
&\psi^{*(l)(m)(n)}(r) = \int_{\infty_2}^{\infty_1} G_\psi(r,r') j_{\psi^*}^{(l)(m)(n)}(r') \, dr' , \label{alphai_psi_lmn}
\end{align}
where $G_\alpha$ and $G_\psi$ are Green's functions obeying
\begin{align}
\Box_i G_\alpha(r,r') = \delta(r-r'), \qquad \qquad \Box_\psi G_\psi = \delta(r-r').
\end{align}
The Green's functions are
\begin{align}
&G_i(r,r') = \frac{1}{r^{\prime3}f(r')W_X(r')} \left[ \Theta(r-r') X_1(r) X_2(r') + \Theta(r' - r) X_2(r) X_1(r') \right], \nonumber \\
&G_\psi(r,r') = \frac{1}{r^{\prime5}f(r')W_Y(r')} \left[ \Theta(r-r') Y_1(r) Y_2(r') + \Theta(r' - r) Y_2(r) Y_1(r') \right].
\end{align}
where the function $\Theta(r-r')$ is a step function compatible with the radial contour. $X_{1,2}$ and $Y_{1,2}$ are fundamental solutions to equations $\Box_i X =0$ and $\Box_\psi Y =0$, respectively,
\begin{align}
&X_1 (r) = {\rm i} \pi - \frac{1}{2} \log \frac{r^2 - r_h^2}{r^2 + r_h^2}, \qquad \qquad \qquad \qquad \qquad \, \, \, \, X_2(r) = - \frac{1}{2} \log \frac{r^2 - r_h^2}{r^2 + r_h^2}, \nonumber \\
&Y_1(r) = - \frac{2{\rm i \pi}}{r^2 + r_h^2} - \frac{\log r - \log (r^2 - r_h^2)}{r^2 + r_h^2}, \qquad \qquad Y_2(r) = - \frac{\log r - \log (r^2 - r_h^2)}{r^2 + r_h^2}.
\end{align}
For convenience, the integration constants in the fundamental solutions above have been determined according to the boundary conditions \eqref{AdS_condition_perturb}. Finally, $W_X$ and $W_Y$ are the Wronskian determinants of fundamental solutions
\begin{align}
& W_X(r) \equiv X_2(r) \partial_r X_1(r) - X_1(r) \partial_r X_2(r) = \frac{2 {\rm i} \pi r_h^2}{r^3 f(r)},\nonumber \\
& W_Y(r) \equiv Y_2(r) \partial_r Y_1(r) - Y_1(r) \partial_r Y_2(r) = \frac{2 {\rm i} \pi r_h^4}{r^5 f(r)}.
\end{align}

\subsection{Holographic results}

With perturbative solutions obtained, it is straightforward (although tedious) to calculate the total bulk action \eqref{Seff_bulk}. We defer the details to appendix \ref{holo_calculation}. Here, we would like to stress that, as shown in appendix \ref{holo_calculation}, the calculations by holographic Schwinger-Keldysh do exactly yield the EFT action of section \ref{EFT_construction}, particularly confirming the proposal of various symmetries. Thus, our study directly demonstrates that, Model F of \cite{RevModPhys.49.435} provides a leading order approximation for holographic superfluid near the critical point.

We advance by summarizing holographic prediction for various coefficients in the effective Lagrangian. For the diffusive part $\mathcal L_{diff}$ \eqref{L_diff}, we obtain\footnote{Recall that we have set $r_h =1$ so that $\pi T =1$.}
\begin{align}
& a_0 = 2, \qquad a_1 = 0, \qquad a_2 = -1, \qquad a_3 = 0, \qquad a_4 =0, \qquad  a_5 = \log2 ,\nonumber \\
& a_6 = - \frac{\log2}{2}, \qquad a_7 = -\frac{\log2}{2}, \qquad a_8 = \frac{\log2}{2}, \qquad a_9 = 0, \qquad u_0 = 0, \nonumber \\
& u_1= \frac{1}{\pi}, \qquad u_2 =0, \qquad u_3 =0, \qquad u_4 =0, \qquad u_5 = \frac{\pi}{24}, \qquad u_6 = - \frac{\pi}{8}, \nonumber \\
& u_7= \frac{\pi}{16}, \qquad u_8 = - \frac{\pi}{16}, \label{a_u_value}
\end{align}
which are in perfect agreement with relevant results of \cite{deBoer:2018qqm,Bu:2020jfo,Baggioli:2023tlc} obtained via different techniques. Here, we stress that the results $a_1 = a_4 = a_9 = u_3 = u_4 =0$ are specific to the holographic model. Moreover, as discussed in \cite{Baggioli:2023tlc}, the values of $a_1$ and $a_4$ seem to be frame-dependent \cite{Glorioso:2017fpd}, whose further exploration is left as a future task.

For the order parameter part $\mathcal L_{\Delta}$ \eqref{L_Delta}, the holographic model predicts
\begin{align}
& b_0 = \frac{1}{2} \delta \mu, \qquad b_1 = - \frac{1}{4}(1-3 {\rm i}), \qquad b_2 = \frac{1+2\log2}{8} + {\rm i} \frac{\log2}{8}, \qquad b_3 = - \frac{1}{4}, \nonumber \\
& v_0 = \frac{1}{2\pi} - \frac{\log2}{2\pi} \delta\mu, \qquad v_1 = \frac{\log2}{4\pi}. \qquad v_2= - 0.25775, \qquad v_3 = 0, \label{b_v_value}
\end{align}
where $\delta\mu = \mu - \mu_0 \simeq (T_c - T)$ with $T_c$ the critical temperature. Obviously, the coefficient $b_0$ would vanish when the system is exactly on the critical point. The coefficient $b_1$ is complex, which is different from that of weakly coupled theory \cite{kamenev_2023}. This may result in interesting phenomena. From the mode analysis, see \eqref{D_Gamma}, the ratio $b_0/{\rm Re}(b_1)$ determines the relaxation rate for the order parameter, which approaches zero as $T \to T_c$. Similar to $b_0$ and $v_0$, we expect that $v_3$ will receive $\delta\mu$-corrections, which is inspired by the study of \cite{Baggioli:2023tlc}.

For the interaction part $\mathcal L_{int}$ \eqref{L_int}, the holographic results are
\begin{align}
& c_0 = c_1 = \frac{1}{2}, \qquad c_2 = \frac{1}{4}, \qquad  c_3 = 0.0208333, \qquad 2c_4 = c_5 = -0.346573, \nonumber \\
& w_0 = - \frac{\log2}{4\pi}, \qquad w_1 = - \frac{\log2}{2\pi}, \qquad w_2 = - \frac{1}{8\pi}, \qquad w_3 = - \frac{1}{8\pi}, \nonumber \\
& w_4 = 0.000263406, \qquad w_5 = 0.00105363, \qquad w_6 = - \frac{1}{4\pi}, \qquad w_7 = 0.0900764, \nonumber \\
& w_8 = 0.090075, \qquad w_9 = 0.0191166, \qquad w_{10} = \frac{11}{96} - \frac{\log^22}{8\pi^2}, \qquad w_{11}= \frac{7}{192}, \nonumber \\
& w_{12}= -0.00466688, \qquad w_{13}= -0.0342099, \qquad w_{14} = -0.0713795, \nonumber \\
& w_{15} = -0.0376026, \qquad w_{16} = 0.000129006, \qquad w_{17} = - 0.00312451, \nonumber \\
& w_{18} = 0.0160281. \label{c_w_value}
\end{align}
The coefficients $c_3$, $w_4$, $w_5$ and $w_{12}$ were previously obtained in \cite{Bu:2021clf}. Due to high nonlinearity, we are able to obtain partial analytical results. Nonetheless, the results \eqref{c_w_value} satisfy all the symmetry constraints of section \ref{EFT_construction}, which can be viewed as a nontrivial support for our calculation. Here, non-Gaussianity, including not only nonlinear interactions between $r$-variables and noises but also nonlinear interactions among noises, is introduced in a systematic way. The phenomenological consequences would be explored using the trick of \cite{Lin:2023bli}.

\section{Summary and Outlook} \label{Summary_Outlook}

We formulated a Schwinger-Keldysh EFT for a nearly critical superfluid system when the temperature is slightly above a critical value. One dynamical mode in the EFT corresponds to the conserved U(1) charge. In addition, given the phenomenon of critical slowing down, non-conserved order parameter was also retained in such an EFT. Therefore, the effective theory we constructed describes dynamics of two scalar fields: a neutral scalar for the conserved U(1) charge and a complex scalar for the non-conserved order parameter.

The EFT Lagrangian is stringently constrained by a set of symmetries. Among others, two of them are worth emphasizing. One is the dynamical KMS symmetry, which originates from time-reversal invariance of the microscopic system and relies on (local) thermal equilibrium. In general, such a symmetry relates terms with different number of time-derivatives but equal number of fields. The other one is the chemical shift symmetry, which ties terms with different number of fields and thus provides a systematic way of generating interactions in the EFT.

Through the holographic Schwinger-Keldysh technique, we derived the EFT Lagrangian of a critical holographic superfluid model. It turns out that holographic derivation perfectly matches the EFT constructed based on symmetry principles. Moreover, holographic calculation also gives values for all Wilsonian coefficients in the EFT Lagrangian.

The studies conducted in present work, both EFT construction via symmetries and holographic calculations, significantly extended relevant results in the literature \cite{RevModPhys.49.435,Bu:2021clf,Donos:2023ibv}. This is mainly reflected on the treatment on thermal fluctuations: not only white noises but also non-Gaussian ones were accounted for systematically. Their phenomenological effects could be explored along the line of \cite{Lin:2023bli}.

The present work can be extended in several directions. First, one could study superfluid EFT in low temperature phase \cite{Yin:2013fwa,Landry:2020ire,Jeong:2023las}. Then, the order parameter gains a background, rendering the chemical shift symmetry to be abandoned. Here, it is of interest to explore symmetry breaking patterns from perspectives of both gravity and boundary EFT. Among others, such a study would give rise to an effective model governing the evolution of order parameter, in the form of Gross–Pitaevskii equation \cite{kamenev_2023}, supporting numerical simulations performed recently in \cite{Xia:2021jzh,Yan:2022jfc,Wittmer:2020mnm,Yang:2022foe,Xia:2019eje}. Second, one would consider gravitational backreaction in the bulk. This corresponds to including extra gapless modes associated with energy and momentum in the boundary EFT \cite{Crossley:2015evo}. Last but not the least, it is worth exploring EFT-inspired improvement over stochastic models used in the study of dynamical critical phenomena \cite{RevModPhys.49.435}. We hope to study these projects in the near future.

\appendix

\section{Source terms} \label{source_terms}

In this appendix, we collect the source terms. In accord with the $\lambda$-expansion, i.e., expansion in number of boundary fields, we classify the source terms into different categories. Within each category, we further group source terms by number of boundary derivatives or by $\delta\mu$-expansion.

$\bullet$ Source terms linear in $\lambda$

In this category, at the lowest order $\mathcal{O}(\xi^0 \lambda^1 \kappa^0)$, all the source terms vanish
\begin{align}
j_v^{(0)(1)(0)}(r) = j_i^{(0)(1)(0)}(r) = j_\psi^{(0)(1)(0)}(r) = j_{\psi^*}^{(0)(1)(0)}(r) = 0. \label{source_010}
\end{align}
The next order corresponds to a $\delta\mu$ correction, i.e., $\mathcal{O}(\xi^0 \lambda^1 \kappa^1)$. The source terms are
\begin{align}
& j_v^{(0)(1)(1)}(r) = j_i^{(0)(1)(1)}(r) = 0, \nonumber \\
& j_\psi^{(0)(1)(1)}(r) = - \frac{2r }{f(r)} \phi_0(r) \delta \phi_0(r) \psi^{(0)(1)(0)}(r), \nonumber \\
& j_{\psi^*}^{(0)(1)(1)}(r) = - \frac{2r}{f(r)} \phi_0(r) \delta \phi_0(r) \psi^{*(0)(1)(0)}(r), \label{source_011}
\end{align}
where $\delta \phi_0 = \delta \mu (1-1/r^2)$. Hereafter, we will ignore $\delta\mu$-correction. Then, at the next order $\mathcal O (\xi^1 \lambda^1 \kappa^0)$, we have
\begin{align}
    &j_v^{(1)(1)(0)}(r) = - \frac{2r}{f(r)} \partial_r \partial_v \alpha_v^{(0)(1)(0)}(r) - \left[ \frac{1}{f(r)} - \frac{r \partial_r f(r)}{f^2(r)} \right] \partial_v \alpha_v^{(0)(1)(0)}(r) , \nonumber \\
    &j_i^{(1)(1)(0)}(r) = - 2r \partial_r \partial_v \alpha_i^{(0)(1)(0)} (r) - \partial_v \alpha_i^{(0)(1)(0)}(r), \nonumber \\
    &j_\psi^{(1)(1)(0)}(r) = -2r^3 \partial_r \partial_v \psi^{(0)(1)(0)} (r) - 3r^2 \partial_v \psi^{(0)(1)(0)} (r) - \frac{2 {\rm i} r}{f(r)} \phi_0 \partial_v \psi^{(0)(1)(0)} (r), \nonumber \\
    &j_{\psi^*}^{(1)(1)(0)}(r) = -2r^3 \partial_r \partial_v \psi^{*(0)(1)(0)}(r) - 3r^2 \partial_v \psi^{*(0)(1)(0)}(r) + \frac{2{\rm i}r}{f(r)} \phi_0 \partial_v \psi^{*(0)(1)(0)} (r).  \label{source_110}
\end{align}
At the order $\mathcal O (\xi^2 \lambda^1 \kappa^0)$, we have
\begin{align}
    j_v^{(2)(1)(0)}(r) = & - \frac{2r}{f(r)} \partial_r \partial_v \alpha_v^{(1)(1)(0)}(r) - \left[ \frac{1}{f(r)} - \frac{r \partial_r f(r)}{f^2(r)} \right] \partial_v \alpha_v^{(1)(1)(0)}(r) \nonumber \\
    & - \frac{1}{r f^2(r)} \partial_v^2 \alpha_v^{(0)(1)(0)}(r) - \frac{1}{r f(r)} \left[ {\vec\partial}^{\,2} \alpha_v^{(0)(1)(0)}(r) - \partial_v \partial_i \alpha_i^{(0)(1)(0)}(r) \right], \nonumber \\
    j_i^{(2)(1)(0)}(r) = & - 2r \partial_r \partial_v \alpha_i^{(1)(1)(0)}(r) - \partial_v \alpha_i^{(1)(1)(0)}(r) - \frac{1}{r f(r)} \partial_v \partial_i \alpha_v^{(0)(1)(0)}(r) \nonumber \\
    & - \frac{1}{r} \left[ {\vec\partial}^{\,2} \alpha_i^{(0)(1)(0)}(r) - \partial_i \partial_k \alpha_k^{(0)(1)(0)}(r) \right], \nonumber \\
    j_\psi^{(2)(1)(0)}(r) = & -2r^3 \partial_r \partial_v \psi^{(1)(1)(0)}(r) - 3r^2 \partial_v \psi^{(1)(1)(0)}(r) - r {\vec\partial}^{\,2} \psi^{(0)(1)(0)}(r) \nonumber \\
    & - \frac{2{\rm i} r}{f(r)} \phi_0 \partial_v \psi^{(1)(1)(0)}(r), \nonumber \\
    j_{\psi^*}^{(2)(1)(0)}(r) = & -2r^3 \partial_r \partial_v \psi^{*(1)(1)(0)}(r) - 3r^2 \partial_v \psi^{*(1)(1)(0)}(r) - r {\vec \partial}^{\,2} \psi^{*(0)(1)(0)}(r) \nonumber \\
    & + \frac{2{\rm i} r}{f(r)} \phi_0 \partial_v \psi^{*(1)(1)(0)}(r). \label{source_210}
\end{align}

$\bullet$ Source terms quadratic in $\lambda$

Notice that in this category, we will not consider $\delta\mu$-correction. At the lowest order $\mathcal O (\xi^0 \lambda^2 \kappa^0)$, we have the source terms
\begin{align}
    &j_v^{(0)(2)(0)}(r) = \frac{2r}{f(r)} \phi_0 \psi^{(0)(1)(0)}(r) \psi^{*(0)(1)(0)}(r), \nonumber \\
    &j_i^{(0)(2)(0)}(r) =0, \nonumber \\
    &j_\psi^{(0)(2)(0)}(r) = - \frac{2r}{f(r)}\phi_0 \alpha_v^{(0)(1)(0)}(r) \psi^{(0)(1)(0)}(r), \nonumber \\
    &j_{\psi^*}^{(0)(2)(0)}(r) = - \frac{2r}{f(r)} \phi_0 \alpha_v^{(0)(1)(0)}(r) \psi^{*(0)(1)(0)}(r).  \label{source_020}
\end{align}
The leading derivative correction is at the order $\mathcal O(\xi^1 \lambda^2 \kappa^0)$. The relevant source terms are
\begin{align}
    j_v^{(1)(2)(0)} = & - \frac{2r}{f(r)} \partial_r \partial_v \alpha_v^{(0)(2)(0)}(r) - \left[ \frac{1}{f(r)} - \frac{r \partial_r f(r)}{f^2(r)} \right] \partial_v \alpha_v^{(2)(0)}(r) \nonumber \\
    & + \frac{{\rm i}r}{f(r)} \left[ \psi^{*(0)(1)(0)}(r) \partial_v \psi^{(0)(1)(0)}(r) - \psi^{(0)(1)(0)}(r) \partial_v \psi^{*(0)(1)(0)}(r) \right] \nonumber \\
    & + \frac{2r \phi_0}{f(r)} \left[ \psi^{(0)(1)(0)}(r) \psi^{*(1)(1)(0)}(r) + \psi^{(1)(1)(0)}(r) \psi^{*(0)(1)(0)}(r) \right], \nonumber \\
    j_i^{(1)(2)(0)}(r) = & - 2r \partial_r \partial_v \alpha_i^{(2)(0)}(r) - \partial_v \alpha_i^{(2)(0)}(r) + {\rm i}r \left[ \psi^{*(0)(1)(0)}(r) \partial_i \psi^{(0)(1)(0)}(r) \right. \nonumber \\
    & \left. - \psi^{(0)(1)(0)}(r) \partial_i \psi^{*(0)(1)(0)}(r) \right], \nonumber \\
    j_\psi^{(1)(2)(0)}(r) = & -2r^3 \partial_r \partial_v \psi^{(0)(2)(0)}(r) - 3 r^2 \partial_v \psi^{(0)(2)(0)}(r) - \frac{2{\rm i}r}{f(r)} \phi_0 \partial_v \psi^{(0)(2)(0)}(r) \nonumber \\
    & - \frac{{\rm i}r}{f(r)} \psi^{(0)(1)(0)}(r) \partial_v \alpha_v^{(0)(1)(0)}(r) + 2{\rm i}r \alpha_i^{(0)(1)(0)} (r)\partial_i \psi^{(0)(1)(0)}(r) \nonumber \\
    & + {\rm i}r \psi^{(0)(1)(0)}(r) \partial_i \alpha_i^{(0)(1)(0)}(r) - \frac{2{\rm i}r}{f(r)} \alpha_v^{(0)(1)(0)}(r) \partial_v \psi^{(0)(1)(0)}(r)
    \nonumber \\
    & - \frac{2r}{f(r)} \phi_0 \left[ \alpha_v^{(0)(1)(0)}(r) \psi^{(1)(1)(0)}(r) + \alpha_v^{(1)(1)(0)}(r) \psi^{(0)(1)(0)}(r) \right], \nonumber \\
    j_{\psi^*}^{(1)(2)(0)}(r) = &  -2r^3 \partial_r \partial_v \psi^{*(0)(2)(0)}(r) - 3 r^2 \partial_v \psi^{*(0)(2)(0)}(r) + \frac{2{\rm i} r}{f(r)} \phi_0 \partial_v \psi^{*(0)(2)(0)}(r) \nonumber \\
    & + \frac{{\rm i} r}{f(r)} \psi^{*(0)(1)(0)}(r) \partial_v \alpha_v^{(0)(1)(0)}(r) -  2{\rm i} r \alpha_i^{(0)(1)(0)}(r) \partial_i \psi^{*(0)(1)(0)}(r) \nonumber \\
    & - {\rm i} r \psi^{*(0)(1)(0)}(r) \partial_i \alpha_i^{(0)(1)(0)}(r) + \frac{2{\rm i} r}{f(r)} \alpha_v^{(0)(1)(0)}(r) \partial_v \psi^{*(0)(1)(0)}(r)  \nonumber \\
    & - \frac{2r}{f(r)} \phi_0 \left[ \alpha_v^{(0)(1)(0)}(r) \psi^{*(1)(1)(0)}(r) + \alpha_v^{(1)(1)(0)}(r) \psi^{*(0)(1)(0)}(r) \right].  \label{source_120}
\end{align}
Practically, at this order, we will capture spatial derivatives but ignore time derivatives. This is partially motivated by the scaling assumption $\partial_0 \sim \partial_i^2$.

$\bullet$ Source terms cubic in $\lambda$

In this category, we only consider the leading term corresponding to the order $\mathcal O (\xi^0 \lambda^3 \kappa^0)$. The source terms are
\begin{align}
    j_v^{(0)(3)(0)}(r) = &\frac{2r}{f(r)} \phi_0 \left[ \psi^{(0)(1)(0)}(r) \psi^{*(0)(2)(0)}(r) + \psi^{(0)(2)(0)}(r) \psi^{*(0)(1)(0)}(r) \right] \nonumber \\
    & + \frac{2r}{f(r)} \alpha_v^{(0)(1)(0)}(r) \psi^{(0)(1)(0)}(r) \psi^{*(0)(1)(0)}(r), \nonumber \\
    j_i^{(0)(3)(0)}(r) = & 2r \alpha_i^{(0)(1)(0)}(r) \psi^{(0)(1)(0)}(r) \psi^{*(0)(1)(0)}(r), \nonumber \\
    j_\psi^{(0)(3)(0)}(r) = & - \frac{2r}{f(r)} \phi_0 \left[ \alpha_v^{(0)(1)(0)}(r) \psi^{(0)(2)(0)}(r) + \alpha_v^{(0)(2)(0)}(r) \psi^{(0)(1)(0)}(r) \right] \nonumber \\
    & - \frac{r}{f(r)} \left( \alpha_v^{(0)(1)(0)}(r) \right)^2 \psi^{(0)(1)(0)}(r) + r \left( \alpha_i^{(0)(1)(0)}(r) \right)^2 \psi^{(0)(1)(0)}(r), \nonumber \\
    j_{\psi^*}^{(0)(3)(0)}(r) =&  - \frac{2r}{f(r)} \phi_0 \left[ \alpha_v^{(0)(1)(0)}(r) \psi^{*(0)(2)(0)}(r) + \alpha_v^{(0)(2)(0)}(r) \psi^{*(0)(1)(0)}(r) \right] \nonumber \\
    & - \frac{r}{f(r)} \left( \alpha_v^{(0)(1)(0)}(r) \right)^2 \psi^{*(0)(1)(0)}(r) + r \left( \alpha_i^{(0)(1)(0)}(r) \right)^2 \psi^{*(0)(1)(0)}(r).  \label{source_030}
\end{align}

\section{Details of holographic calculation} \label{holo_calculation}

In this appendix, we present more details on holographic calculation. Particularly, we will record perturbative solutions at each order and compute total bulk action \eqref{Seff_bulk} in details. This will yield the EFT Lagrangian \eqref{L_diff}, \eqref{L_Delta} and \eqref{L_int} (and holographic results for various coefficients as well). In accord with the triple expansion for bulk fields, we expand the total bulk action \eqref{Seff_bulk} similarly
\begin{align}
\mathcal L_{eff} = \sum_{l,m,n} \mathcal L_{eff}^{(l)(m)(n)}.
\end{align}
The following presentation will be in parallel with that of the source terms in appendix \ref{source_terms}.

$\bullet$ Linearized bulk perturbations and Gaussian EFT

First, we consider bulk perturbations linear in boundary fields. Recall that at the lowest order, all the source terms \eqref{source_010} vanish. Then, analytical solutions at this order are
\begin{align}
&\alpha_v^{(0)(1)(0)}(r) = B_{2v}\left( 1- \frac{r_h^2}{r^2} \right), \qquad  r \in [r_h - \epsilon, \infty_2), \nonumber \\
&\alpha_v^{(0)(1)(0)}(r) = B_{1v}\left( 1- \frac{r_h^2}{r^2} \right), \qquad  r \in [r_h - \epsilon, \infty_1), \nonumber \\
&\alpha_i^{(0)(1)(0)}(r) = B_{2i} + B_{ai} \log \frac{r^2-r_h^2}{r^2 + r_h^2}, \nonumber \\
&\psi^{(0)(1)(0)} (r) = \frac{\Delta_2}{r^2 + r_h^2} - \frac{\Delta_a}{2 {\rm i} \pi} \frac{\log r - \log(r^2 - r_h^2)}{r^2 + r_h^2}, \nonumber \\
&\psi^{*(0)(1)(0)} (r) = \frac{\Delta_2^*}{r^2 + r_h^2} - \frac{\Delta_a^*}{2 {\rm i} \pi} \frac{\log r - \log(r^2 - r_h^2)}{r^2 + r_h^2}.
\end{align}
Here, we have imposed all the boundary conditions. From the solutions, we read off results for $J_\mu$, $\psi_b$ and $\psi_b^*$ (cf. \eqref{near_bounday_gauge_fixed})
\begin{align}
& J_{1v}^{(0)(1)(0)} = - B_{1v},  \qquad  J_{2v}^{(0)(1)(0)} = - B_{2v}, \qquad J_{1i}^{(0)(1)(0)} = J_{2i}^{(0)(1)(0)} = - \frac{{\rm i}}{\pi} B_{ai}, \nonumber \\
& \psi_{b1}^{(0)(1)(0)} = \psi_{b2}^{(0)(1)(0)} = - \frac{{\rm i}}{2\pi} \Delta_a, \qquad \psi_{b1}^{^*(0)(1)(0)} = \psi_{b2}^{^*(0)(1)(0)} = - \frac{{\rm i}}{2\pi} \Delta_a^*. \label{J_psib_010}
\end{align}
Via \eqref{Seff_bulk}, the solutions at this order give the following part of EFT Lagrangian
\begin{align}
\mathcal L_{eff}^{(0)(2)(0)} = 2B_{av} B_{rv} + \frac{\rm i }{\pi} B_{ai} ^2+ \frac{\rm i}{2\pi} \Delta_a^* \Delta_a. \label{Leff_020}
\end{align}

Obviously, at the critical point, the coefficient $b_0$ of \eqref{L_Delta} vanishes. In order to account for small deviation from critical point, we proceed by considering $\delta\mu$ correction, i.e., perturbative solutions at the order $\mathcal{O}(\xi^0 \lambda^1 \kappa^1 )$. Plugging the source terms \eqref{source_011} into \eqref{alphav_lmn} and \eqref{alphai_psi_lmn}, it is straightforward to obtain
\begin{align}
& J_{1v}^{(0)(1)(1)} = J_{2v}^{(0)(1)(1)} = J_{1i}^{(0)(1)(1)} = J_{2i}^{(0)(1)(1)} =0, \nonumber \\
& \psi_{b1}^{(0)(1)(1)} = \delta \mu \left[-\frac{1}{4} \Delta_a + \frac{{\rm i} \log2}{2 \pi} \Delta_a - \frac{1}{2} \Delta_r \right], \nonumber \\
& \psi_{b2}^{(0)(1)(1)} = \delta \mu \left[\frac{1}{4} \Delta_a + \frac{{\rm i} \log2}{2 \pi} \Delta_a - \frac{1}{2} \Delta_r \right], \nonumber \\
& \psi_{b1}^{*(0)(1)(1)} = \delta \mu \left[ -\frac{1}{4} \Delta_a^* + \frac{{\rm i} \log2}{2\pi} \Delta_a^* - \frac{1}{2} \Delta_r^* \right], \nonumber \\
& \psi_{b2}^{*(0)(1)(1)} = \delta \mu  \left[ \frac{1}{4} \Delta_a^* + \frac{ {\rm i} \log2}{2\pi} \Delta_a^* - \frac{1}{2} \Delta_r^* \right]. \label{J_psib_011}
\end{align}
Here, due to logarithmic singularities near the horizon, we have computed the radial integral by splitting the radial contour
\begin{align}
\int_{\infty_2}^{\infty_1} dr \cdots = \int_{\infty_2}^{r_h +\epsilon} dr \cdots + \int_{\mathcal C} dr \cdots + \int_{r_h +\epsilon}^{\infty_1} dr \cdots, \label{contour_split}
\end{align}
where the integral along the infinitesimal circle $\mathcal C$ will be calculated in the polar coordinate. Interestingly, the first part and third part in \eqref{contour_split} will cancel significantly. Actually, this treatment will be employed in the calculations of higher order perturbations and the bulk part of \eqref{Seff_bulk}. Plugging \eqref{J_psib_011} into \eqref{Seff_bulk}, we obtain
\begin{align}
\mathcal L_{eff}^{(0)(2)(1)} = \delta\mu \left\{ \frac{1}{2}\Delta_a^* \Delta_r + \frac{1}{2}\Delta_a \Delta_r^* -\frac{{\rm i} \log2}{2\pi} \Delta_a^* \Delta_a \right\}. \label{Leff_021}
\end{align}
Clearly, beyond critical point, the coefficient $b_0$ of \eqref{L_Delta} will no longer vanish.

We turn to the order $\mathcal O (\xi^1 \lambda^1 \kappa^0)$. The relevant equations could be solved analytically. However, the expressions for $\psi^{(1)(1)(0)}$ and $\psi^{*(1)(1)(0)}$ are too lengthy to be written here. For later convenience, we record the results for $\alpha_\mu^{(1)(1)(0)}$
\begin{align}
\alpha_{sv}^{(1)(1)(0)}(r) = &\frac{\partial_v B_{sv}}{4r_h} \left(1- \frac{r_h^2}{r^2} \right) \left[ \pi -2\arctan\left( \frac{r}{r_h}\right) + \log\frac{r+r_h}{r-r_h} \right], \nonumber \\
\alpha_i^{(1)(1)(0)}(r) = &\frac{\partial_v B_{2i}}{4r_h} \left[ \pi - 2\arctan\left( \frac{r}{r_h} \right) +2 \log(r+r_h) - \log(r^2 +r_h^2) \right] \nonumber \\
&- \frac{\partial_v B_{ai}}{8\pi r_h} \left[ -(2- {\rm i}) \pi - 2 {\rm i} \arctan\left( \frac{r}{r_h} \right) -{\rm i} \log \frac{r-r_h}{r+r_h} \right] \log \frac{r^2 - r_h^2}{r^2 + r_h^2}.
\end{align}
From the solutions, we read off $J_\mu$ and $\psi_b$
\begin{align}
& J_{1v} ^{(1)(1)(0))} = J_{2v} ^{(1)(1)(0))} = 0, \nonumber \\
& J_{1i}^{(1)(1)(0))} = \frac{1}{4} \partial_v B_{ai} - \frac{1}{2}\partial_v B_{ri}, \qquad J_{2i} ^{(1)(1)(0))} = -\frac{1}{4} \partial_v B_{ai} - \frac{1}{2} \partial_v B_{ri}, \nonumber \\
& \psi_{b2}^{(1)(1)(0)} = \frac{1}{8}(1 + 3 {\rm i}) \partial_v \Delta_a - \frac{\log2}{4\pi} \partial_v \Delta_a + \frac{1}{4} (1 - 3{\rm i}) \partial_v \Delta_r, \nonumber \\
& \psi_{b1}^{(1)(1)(0)} = - \frac{1}{8}(1+ 3 {\rm i}) \partial_v \Delta_a - \frac{\log2} {4\pi} \partial_v\Delta_a + \frac{1}{4} (1 - 3{\rm i}) \partial_v \Delta_r, \nonumber \\
& \psi_{b2}^{*(1)(1)(0)} = \frac{1}{8}(1 - 3{\rm i}) \partial_v\Delta_a^* + \frac{\log2}{4\pi} \partial_v \Delta_a^* + \frac{1}{4} (1 + 3{\rm i}) \partial_v\Delta_r^*, \nonumber \\
& \psi_{b1}^{*(1)(1)(0)} = -\frac{1}{8} (1 - 3{\rm i}) \partial_v\Delta_a^* + \frac{\log2}{4\pi}  \partial_v \Delta_a^* + \frac{1}{4} (1 + 3 {\rm i}) \partial_v\Delta_r^*.
\end{align}
Immediately, from \eqref{Seff_bulk} we obtain the following action
\begin{align}
\mathcal L_{eff}^{(1)(2)(0)} = - B_{ai} \partial_v B_{ri} - \frac{1}{4} (1 + 3{\rm i}) \Delta_a\partial_v \Delta_r - \frac{1}{4}(1 - 3{\rm i})\Delta_a^*\partial_v \Delta_r + \frac{ \log2}{4 \pi} \Delta_a^*  \partial_v \Delta_a. \label{Leff_120}
\end{align}

We would also like to work out solutions at the order $\mathcal{O}(\xi^2 \lambda^1 \kappa^0)$. Based on general formulas \eqref{alphav_lmn} and \eqref{alphai_psi_lmn}, we are able to obtain
\begin{align}
J_{2v}^{(2)(1)(0))} = & \left( \frac{1}{4} - \frac{\log2}{2} \right)\partial_i^2 B_{2v} + \left(\frac{1}{8} - \frac{{\rm i} \pi}{48} - \frac{\log2}{8} \right) \partial_v \partial_i B_{ai} \nonumber \\
& - \left(\frac{1}{4} - \frac{\log2}{4} \right) \partial_v \partial_i B_{ri} + \frac{1}{2} \partial_v^2 B_{2v}, \nonumber \\
J_{1v}^{(2)(1)(0))} = &\left( \frac{1}{4} -\frac{\log2}{2} \right) \partial_i^2 B_{1v} - \left( \frac{1}{8} + \frac{{\rm i} \pi}{48} - \frac{\log2}{8} \right) \partial_v \partial_i B_{ai} \nonumber \\
&  - \left( \frac{1}{4} - \frac{\log2}{4} \right) \partial_v \partial_i B_{ri} + \frac{1}{2} \partial_v^2 B_{1v}, \nonumber \\
J_{2i}^{(2)(1)(0))} =& -\left(\frac{1}{8} + \frac{{\rm i}\pi}{8} + \frac{\log2}{8} \right) \partial_v^2 B_{ai} - \left( \frac{1}{8} - \frac{{\rm i}\pi}{16} \right) \left( \partial_k^2 B_{ai} - \partial_i \partial_k B_{ak} \right)\nonumber \\
& + \frac{1}{4}\left( \partial_k^2 B_{ri} - \partial_i \partial_k B_{rk} \right) + \left(\frac{1}{4} - \frac{{\rm i}\pi}{48} - \frac{\log2}{4} \right) \partial_v \partial_i B_{2v} \nonumber \\
& + \left( \frac{1}{4} + \frac{\log2}{4} \right) \partial_v^2 B_{ri} + \frac{{\rm i}\pi}{48} \partial_v \partial_i B_{1v}, \nonumber \\
J_{1i}^{(2)(1)(0))} =& \left( \frac{1}{8} - \frac{{\rm i}\pi}{8} + \frac{\log2}{8} \right) \partial_v^2 B_{ai} + \left( \frac{1}{8} + \frac{{\rm i}\pi}{16} \right) \left( \partial_k^2 B_{ai} - \partial_i \partial_k B_{ak}\right) \nonumber \\
&+ \frac{1}{4} \partial_k^2 B_{ri} - \frac{1}{4} \partial_i \partial_k B_{rk} + \left(\frac{1}{4} + \frac{{\rm i}\pi}{48} - \frac{\log2}{4} \right) \partial_v \partial_i B_{2v} \nonumber \\
&+ \left( \frac{1}{4} + \frac{\log2}{4} \right) \partial_v^2 B_{ri} + \frac{{\rm i}\pi}{48} \partial_v \partial_i B_{1v},  \nonumber \\
\psi_{b2}^{(2)(1)(0)} =& (0.149143 + 0.214428{\rm i}) \partial_v^2 \Delta_a - \left( \frac{1}{8} + \frac{(2+{\rm i}) \log2}{8} \right) \partial_v^2 \Delta_r \nonumber \\
&+ \frac{1}{8} \partial_i^2 \Delta_a - \frac{1}{4} \partial_i^2 \Delta_r, \nonumber \\
\psi_{b1}^{(2)(1)(0)} =& (-0.149144+0.301071{\rm i}) \partial_v^2 \Delta_a - \left( \frac{1}{8} + \frac{(2+{\rm i}) \log2}{8} \right) \partial_v^2 \Delta_r \nonumber \\
& - \frac{1}{8} \partial_i^2 \Delta_a - \frac{1}{4} \partial_i^2 \Delta_r, \nonumber \\
\psi_{b2}^{*(2)(1)(0)} =& (0.149144+0.301071{\rm i}) \partial_v^2 \Delta_a^* - \left( \frac{1}{8} + \frac{(2-{\rm i}) \log2}{8} \right) \partial_v^2 \Delta_r^* \nonumber \\
&+ \frac{1}{8} \partial_i^2 \Delta_a^* - \frac{1}{4} \partial_i^2 \Delta_r^*, \nonumber \\
\psi_{b1}^{*(2)(1)(0)} =& (-0.149143+0.214428{\rm i}) \partial_v^2 \Delta_a^* - \left( \frac{1}{8} + \frac{(2-{\rm i}) \log2}{8} \right) \partial_v^2 \Delta_r^* \nonumber \\
& - \frac{1}{8} \partial_i^2 \Delta_a^* - \frac{1}{4} \partial_i^2 \Delta_r^*.
\end{align}
Here, we were unable to compute coefficients of $\partial_v^2 \Delta_a^*$ and $\partial_v^2 \Delta_a$ analytically. The calculation of these terms will go through the treatment of \eqref{contour_split}, with each part involving singularity near the horizon. So, we take a tiny value for $\epsilon$ and compute each part of \eqref{contour_split} numerically. We have checked that the final result (i.e., summation of the three parts in \eqref{contour_split}) is insensitive to the specific choice for $\epsilon$.

Then, we obtain second order derivative terms for the EFT Lagrangian
\begin{align}
\mathcal L_{eff}^{(2)(2)(0)} = & \log2\, B_{av} \partial_i ^2 B_{rv} + \frac{{\rm i}\pi}{24} B_{av} \partial_v \partial_i B_{ai} - \frac{\log2}{2} \left( B_{av} \partial_v \partial_i B_{ri} + B_{rv} \partial_v \partial_i B_{ai} \right) \nonumber \\
& - \frac{{\rm i}\pi}{8} B_{ai} \partial_v^2 B_{ai} + \frac{{\rm i}\pi}{16} B_{ai} \partial_k^2 B_{ai} + \frac{\log2}{2} B_{ai} \partial_v^2 B_{ri} - \frac{{\rm i}\pi}{16} B_{ai} \partial_i \partial_k B_{ak} \nonumber \\
& - \frac{1}{2} B_{ai} \partial_v \partial_k B_{rk} + \frac{1}{4} B_{ai} \partial_k^2 B_{rv} + \frac{1}{4} B_{ri} \partial_k^2 B_{av} + \frac{1}{4} \Delta_a^* \partial_i ^2 \Delta_r + \frac{1}{4} \Delta_r^* \partial_i^2 \Delta_a \nonumber \\
&+ \left( \frac{1+ 2\log2}{8} + \frac{{\rm i}\log2}{8} \right) \Delta_a^* \partial_v^2 \Delta_r + \left( \frac{1+ 2\log2}{8} - \frac{{\rm i}\log2}{8} \right) \Delta_r^* \partial_v^2 \Delta_a \nonumber \\
& - 0.25775{\rm i} \Delta_a^* \partial_v^2 \Delta_a.  \label{Leff_220}
\end{align}

So far, we have computed Gaussian part of the EFT action, accurate up to second order in boundary derivatives. The results \eqref{Leff_020}, \eqref{Leff_021}, \eqref{Leff_120} and \eqref{Leff_220} perfectly match \eqref{L_diff} and \eqref{L_Delta}\footnote{The cubic and quartic terms in \eqref{L_Delta}, arising from the covariant derivatives, will be reported later.}, giving rise to holographic results \eqref{a_u_value} and \eqref{b_v_value}.

$\bullet$ Nonlinear bulk perturbations and non-Gaussian EFT

Now, we consider bulk perturbations nonlinear in $\lambda$. The leading terms are of order $\mathcal O (\xi^0 \lambda^2 \kappa^0)$. With the source terms \eqref{source_020} known analytically, we are able to analytically calculate $J_\mu$ and $\psi_b$ using the formulas \eqref{alphav_lmn} and \eqref{alphai_psi_lmn}. The results are
\begin{align}
J_{2v}^{(0)(2)(0))} = & \left(- \frac{11}{192} - \frac{{\rm i}\log2}{8\pi} + \frac{\log^22}{16 \pi^2}\right) \Delta_a \Delta_a^* + \left( \frac{1}{8} + \frac{{\rm i}\log2}{8\pi} \right) \Delta_r \Delta_a^* \nonumber \\
& + \left( \frac{1}{8} + \frac{{\rm i}\log2}{8\pi} \right) \Delta_a \Delta_r^* - \frac{1}{4} \Delta_r \Delta_r^*, \nonumber \\
J_{1v}^{(0)(2)(0))} =&  \left(- \frac{11}{192} + \frac{{\rm i}\log2}{8\pi} + \frac{\log^22}{16 \pi^2}\right) \Delta_a \Delta_a^* + \left( - \frac{1}{8} + \frac{{\rm i} \log(2)}{8\pi} \right) \Delta_r \Delta_a^* \nonumber \\
& + \left(- \frac{1}{8} + \frac{{\rm i}\log2}{8\pi} \right) \Delta_a \Delta_r^* - \frac{1}{4} \Delta_r \Delta_r^*, \nonumber \\
J_{1i}^{(0)(2)(0))} =& J_{2i} ^{(0)(2)(0))} = 0, \nonumber \\
\psi_{b2}^{(0)(2)(0)} =& \left( \frac{1}{96} + \frac{{\rm i}\log2}{8\pi} + \frac{\log^22}{8\pi^2}\right) B_{1v} \Delta_a + \frac{{\rm i}\log2}{4 \pi} B_{1v} \Delta_r \nonumber \\
& + \left( \frac{23}{96} + \frac{3{\rm i} \log2}{8 \pi} - \frac{\log^22}{8 \pi^2} \right) B_{2v} \Delta_a + \left(- \frac{1}{2} - \frac{{\rm i} \log2}{4 \pi} \right) B_{2v} \Delta_r, \nonumber \\
\psi_{b1}^{(0)(2)(0)} =& \left( - \frac{23}{96} + \frac{3{\rm i}\log2}{8 \pi} + \frac{\log^22}{8 \pi^2} \right) B_{1v} \Delta_a + \left( - \frac{1}{2} + \frac{{\rm i} \log2}{4 \pi} \right) B_{1v} \Delta_r \nonumber \\
& + \left( - \frac{1}{96} + \frac{{\rm i}\log2}{8\pi} - \frac{\log^22}{8\pi^2}\right) B_{2v} \Delta_a - \frac{{\rm i} \log2}{4 \pi} B_{2v} \Delta_r, \nonumber \\
\psi_{b2}^{*(0)(2)(0)} =& \left( \frac{1}{96} + \frac{{\rm i} \log2}{8\pi} + \frac{\log^22}{8 \pi^2} \right) B_{1v} \Delta_a^* + \frac{{\rm i} \log2}{4 \pi} B_{1v} \Delta_r^* \nonumber \\
& + \left( \frac{23}{96} + \frac{3{\rm i} \log2}{8 \pi} - \frac{ \log^22}{8 \pi^2} \right) B_{2v} \Delta_a^* + \left(- \frac{1}{2} - \frac{{\rm i} \log2}{4 \pi} \right) B_{2v} \Delta_r^*, \nonumber \\
\psi_{b1}^{*(0)(2)(0)} =& \left( - \frac{23}{96} + \frac{3{\rm i}\log2}{8 \pi} + \frac{\log^22}{8 \pi^2} \right) B_{1v} \Delta_a^* + \left( -\frac{1}{2} + \frac{{\rm i} \log(2)}{4 \pi} \right) B_{1v} \Delta_r^*  \nonumber \\
& + \left(- \frac{1}{96} + \frac{{\rm i}\log2}{8\pi} -
\frac{\log^22}{8\pi^2} \right) B_{2v} \Delta_a^* - \frac{{\rm i} \log2}{4 \pi} B_{2v} \Delta_r^*.
\end{align}
The EFT Lagrangian at the order $\mathcal{O}(\xi^0\lambda^3 \kappa^0)$ is
\begin{align}
\mathcal L_{eff}^{(0)(3)(0)} =& \frac{1}{2} B_{av} \Delta_r \Delta_r^* + \frac{1}{2} B_{rv} \Delta_a \Delta_r^* + \frac{1}{2} B_{rv} \Delta_r \Delta_a^* - \frac{{\rm i}\log2}{4 \pi} B_{av} \Delta_a \Delta_r^* \nonumber \\
&  - \frac{{\rm i} \log2}{4 \pi} B_{av} \Delta_r \Delta_a^* - \frac{{\rm i}\log2}{2 \pi} B_{rv} \Delta_a \Delta_a^* + \left( \frac{11}{96} - \frac{ \log^22}{8 \pi^2}\right) B_{av} \Delta_a \Delta_a^*. \label{Leff_030}
\end{align}

In order to generate first order derivative correction to \eqref{Leff_030}, we calculate bulk perturbations at the order $\mathcal O (\xi^1 \lambda^3 \kappa^0)$. For simplicity, we ignore all time-derivative terms. Then, we are able to obtain $J_\mu$ and $\psi_b$ analytically at this order
\begin{align}
J_{1v}^{(1)(2)(0))} =& = J_{2v}^{(1)(2)(0))} =0, \nonumber \\
J_{2i}^{(1)(2)(0))} =& \left( \frac{{\rm i}}{16} + \frac{1}{16\pi} \right)\left( \Delta_a^* \partial_i \Delta _r + \Delta_r^* \partial_i \Delta_a -\Delta_a \partial_i \Delta_r^* - \Delta_r \partial_i \Delta_a^*  \right) \nonumber \\
& + \frac{7{\rm i}}{384} \left( \Delta_a \partial_i \Delta_a^* - \Delta_a^* \partial_i \Delta_a \right) + \frac{{\rm i}}{8} \left( \Delta_r \partial_i \Delta_r^* -\Delta_r^* \partial_i \Delta_r \right), \nonumber \\
J_{1i}^{(1)(2)(0))} =& \left( \frac{\rm i}{16} - \frac{1}{16\pi} \right) \left(\Delta_a \partial_i \Delta_r^* + \Delta_r \partial_i \Delta_a^* - \Delta_a^* \partial_i \Delta _r - \Delta_r^* \partial_i \Delta_a \right) \nonumber \\
& + \frac{7{\rm i}}{384} \left( \Delta_a \partial_i \Delta_a^* - \Delta_a^* \partial_i \Delta_a \right) + \frac{\rm i}{8} \left( \Delta_r \partial_i \Delta_r^* - \Delta_r^* \partial_i \Delta _r \right), \nonumber \\
\psi_{b2}^{(1)(2)(0)} =& \left( \frac{7\rm i}{192} + \frac{1}{16\pi}\right)
\left( \Delta_a \partial_i B_{ai} + 2 B_{ai} \partial_i \Delta_a \right) - \frac{\rm i}{8} \Delta_a \partial_i B_{ri} + \frac{\rm i}{4} \Delta_r \partial_i B_{ri} \nonumber \\
& - \frac{\rm i}{4} B_{ri} \partial_i \Delta_a  + \frac{\rm i}{2} B_{ri} \partial_i \Delta_r - \left( \frac{\rm i}{8} + \frac{1}{8\pi} \right) \left( \Delta_r \partial_i B_{ai} + 2 B_{ai} \partial_i \Delta_r \right), \nonumber \\
\psi_{b1}^{(1)(2)(0)} =& \left( \frac{7\rm i}{192} - \frac{1}{16\pi}\right) \left(
\Delta_a \partial_i B_{ai} + B_{ai} \partial_i \Delta_a \right) + \frac{\rm i}{8} \Delta_a \partial_i B_{ri} + \frac{\rm i}{4} \Delta_r \partial_i B_{ri} \nonumber \\
& + \frac{\rm i}{4} B_{ri} \partial_i \Delta_a  + \frac{\rm i}{2} B_{ri} \partial_i \Delta_r + \left( \frac{\rm i}{8} - \frac{1}{8\pi} \right) \left( \Delta_r \partial_i B_{ai} + 2 B_{ai} \partial_i \Delta_r \right), \nonumber \\
\psi_{b2}^{*(1)(2)(0)} =& -\left(\frac{7\rm i}{192} + \frac{1}{16\pi} \right) \left(
\Delta_a^* \partial_i B_{ai} + 2 B_{ai} \partial_i \Delta_a^* \right) + \frac{\rm i}{8} \Delta_a^* \partial_i B_{ri} - \frac{\rm i}{4} \Delta_r^* \partial_i B_{ri} \nonumber \\
& + \frac{\rm i}{4} B_{ri} \partial_i \Delta_a^* - \frac{\rm i}{2} B_{ri} \partial_i \Delta_r^* + \left( \frac{\rm i}{8} + \frac{1}{8\pi} \right) \left( \Delta_r^* \partial_i B_{ai} + 2 B_{ai} \partial_i \Delta_r^*\right), \nonumber \\
\psi_{b1}^{*(1)(2)(0)} =& -\left(\frac{7\rm i}{192} - \frac{1}{16\pi} \right) \left(
\Delta_a^* \partial_i B_{ai} + 2B_{ai} \partial_i \Delta_a^* \right)- \frac{\rm i}{8} \Delta_a^* \partial_i B_{ri} - \frac{\rm i}{4} \Delta_r \partial_i B_{ri} \nonumber \\
&- \frac{\rm i}{4} B_{ri} \partial_i \Delta_a^* - \frac{\rm i}{2} B_{ri} \partial_i \Delta_r^* - \left( \frac{\rm i}{8} - \frac{1}{8\pi} \right) \left( \Delta_r^* \partial_i B_{ai} + 2 B_{ai} \partial_i \Delta_r^*\right).
\end{align}
Thus, the EFT Lagrangian at the order $\mathcal{O}(\xi^1 \lambda^3 \kappa^0)$ is
\begin{align}
\mathcal L_{eff}^{(1)(3)(0)} = & - \frac{{\rm i}}{4} B_{ri} \Delta_a^* \partial_i \Delta_r - \frac{\rm i}{4} B_{ri} \Delta_r^* \partial_i \Delta_a + \frac{\rm i}{4} B_{ri} \Delta_a \partial_i \Delta_r^* + \frac{\rm i}{4} B_{ri} \Delta_r \partial_i \Delta_a^* \nonumber \\
& - \frac{{\rm i}}{4} B_{ai} \Delta_r^* \partial_i \Delta_r + \frac{{\rm i}}{4} B_{ai} \Delta_r \partial_i \Delta_r^* + \frac{1}{8\pi} B_{ai} \Delta_a^* \partial_i \Delta_r + \frac{1}{8\pi} B_{ai} \Delta_r^* \partial_i \Delta_a\nonumber \\
& - \frac{1}{8\pi} B_{ai} \Delta_a \partial_i \Delta_r^* - \frac{1}{8\pi} B_{ai} \Delta_r \partial_i \Delta_a^* + \frac{7{\rm i}}{192} B_{ai} \Delta_a \partial_i \Delta_a^* - \frac{7{\rm i}}{192} B_{ai} \Delta_a^* \partial_i \Delta_a. \label{Leff_130}
\end{align}

The last perturbations to be computed are of order $\mathcal O (\xi^0 \lambda^4 \kappa^0)$. With the source terms \eqref{source_030}, we are able to obtain partial analytical results for $J_\mu$ and $\psi_b$:
\begin{align}
J_{2v}^{(0)(3)(0))} =& 0.0064901 {\rm i} B_{1v} \Delta_a \Delta_a^* - 0.00571904 B_{1v} \left( \Delta_a \Delta_r^* + \Delta_r \Delta_a^* \right) \nonumber \\
& - 0.0191166 {\rm i} B_{1v} \Delta_r \Delta_r^* + (0.0356897 + 0.0385473 {\rm i} ) B_{2v} \Delta_a \Delta_a^* \nonumber \\
&- ( 0.0809244 + 0.0450374 {\rm i}) B_{2v} \left( \Delta_a \Delta_r^* + \Delta_r \Delta_a^* \right) \nonumber \\
&+ ( 0.173287 + 0.0191169 {\rm i} ) B_{2v} \Delta_r \Delta_r^*, \nonumber \\
J_{1v}^{(0)(3)(0))} =& - 0.0064901 {\rm i} B_{2v} \Delta_a \Delta_a^* + 0.00571904 B_{2v} \left( \Delta_a \Delta_r^* + \Delta_r \Delta_a^* \right) \nonumber \\
&+ 0.0191166 {\rm i} B_{2v} \Delta_r \Delta_r^* + (0.0356897 - 0.0385473 {\rm i}) B_{1v} \Delta_a \Delta_a^* \nonumber \\
&+ ( 0.0809244 - 0.0450374 {\rm i} ) B_{1v} \left( \Delta_a \Delta_r^* + \Delta_r \Delta_a^* \right)\nonumber \\
&+ ( 0.173287 - 0.0191166 {\rm i} ) B_{1v} \Delta_r \Delta_r^*, \nonumber \\
J_{2i}^{(0)(3)(0))} =& - \frac{7}{192} B_{ri} \Delta_a \Delta_a^* + \frac{8 + 4 {\rm i} \pi - 3 \pi^2}{64 \pi^2} B_{ai} \left( \Delta_r \Delta_a^* + \Delta_a \Delta_r^* \right) \nonumber \\
& + \frac{\pi -{\rm i} }{8 \pi} B_{ri} \left( \Delta_r \Delta_a^* + \Delta_a \Delta_r^* \right)- \left(\frac{\rm i}{4\pi} - \frac{1}{8} \right) B_{ai} \Delta_r \Delta_r^*  \nonumber \\
&- \frac{1}{4} B_{ri} \Delta_r \Delta_r^* + (0.0182292 - 0.00312451 {\rm i}) B_{ai} \Delta_a \Delta_a^* , \nonumber \\
J_{1i}^{(0)(3)(0))} =& - \frac{7}{192} B_{ri} \Delta_a \Delta_a^* + \frac{8 - 4i \pi - 3 \pi^2}{64 \pi^2} B_{ai} \left( \Delta_r \Delta_a^* + \Delta_a \Delta_r^* \right) \nonumber \\
& - \frac{\pi + \rm i}{8 \pi} B_{ri} \left( \Delta_r \Delta_a^* + \Delta_a \Delta_r^* \right) - \left(\frac{\rm i}{4\pi} + \frac{1}{8} \right) B_{ai} \Delta_r \Delta_r^*  \nonumber \\
&- \frac{1}{4} B_{ri} \Delta_r \Delta_r^* - (0.0182292 + 0.00312451 {\rm i} ) B_{ai} \Delta_a \Delta_a^*, \nonumber \\
\psi_{b2}^{(0)(3)(0)} =&- (0.00477255 + 0.0160286 {\rm i} ) B_{1v}^2 \Delta_a \nonumber \\
&- (0.00571908 + 0.0354791 {\rm i}) B_{1v}^2 \Delta_r \nonumber \\
&- (0.0761517 + 0.0610661 {\rm i}) B_{2v}^2 \Delta_a \nonumber \\
&+ ( 0.167568 + 0.0545957 {\rm i}) B_{2v}^2 \Delta_r \nonumber \\
&- ( 0.00571904 + 0.0129802 {\rm i} ) B_{1v} B_{2v} \Delta_a \nonumber \\
&+ ( 0.0114381 - 0.0191166 {\rm i}) B_{1v} B_{2v} \Delta_r \nonumber \\
&- \frac{1}{8} B_{ri}^2 \Delta_a + \frac{1}{4} B_{ri}^2 \Delta_r - \frac{8 + 8 {\rm i} \pi - 3 \pi^2}{64 \pi^2} B_{ai}^2 \Delta_r \nonumber \\
&+ \left(\frac{7}{96} - \frac{\rm i}{8 \pi} \right) B_{ai} B_{ri} \Delta_a - \frac{\pi - {\rm i}}{4 \pi} B_{ai} B_{ri} \Delta_r \nonumber \\
&- (0.0171049 - 0.00312451 {\rm i} ) B_{ai}^2 \Delta_a  \nonumber \\
& + \frac{1}{48} \Delta_r \Delta_r \Delta_r^* - ( 0.0104167 + 0.000526813 {\rm i} ) \Delta_r \Delta_r \Delta_a^* \nonumber \\
&+ (0.00466688 + 0.000263407 {\rm i}) \Delta_a \Delta_a \Delta_r^* \nonumber \\
&+ (0.00933375 + 0.000526813 {\rm i}) \Delta_a \Delta_r \Delta_a^* \nonumber \\
&- (0.0208333 + 0.00105363 {\rm i} ) \Delta_a \Delta_r \Delta_r^* \nonumber \\
&- (0.00233344 + 0.000258912i) \Delta_a \Delta_a \Delta_a^*, \nonumber \\
\psi_{b1}^{(0)(3)(0)} =& (0.0761516 - 0.0610662 {\rm i}) B_{1v}^2 \Delta_a \nonumber \\
&+ ( 0.167568 - 0.0545957 {\rm i}) B_{1v}^2 \Delta_r \nonumber \\
&+ ( 0.00477255 - 0.0160286 {\rm i}) B_{2v}^2 \Delta_a \nonumber \\
&- ( 0.00571908 - 0.0354791 {\rm i}) B_{2v}^2 \Delta_r \nonumber \\
&+ ( 0.00571904 - 0.0129802 {\rm i} ) B_{1v} B_{2v} \Delta_a \nonumber \\
&+ ( 0.0114381 + 0.0191166 {\rm i} ) B_{1v} B_{2v} \Delta_r \nonumber \\
&+ \frac{1}{8} B_{ri}^2 \Delta_a + \frac{1}{4} B_{ri}^2 \Delta_r - \frac{8 - 8 {\rm i} \pi - 3 \pi^2}{64 \pi^2} B_{ai}^2 \Delta_r \nonumber \\
&+ \left(\frac{7}{96} + \frac{\rm i}{8 \pi} \right) B_{ai} B_{ri} \Delta_a + \frac{\pi + {\rm i}}{4 \pi} B_{ai} B_{ri} \Delta_r \nonumber \\
&+ (0.0171049 + 0.00312451 {\rm i}) B_{ai}^2 \Delta_a \nonumber \\
&+ \frac{1}{48} \Delta_r \Delta_r \Delta_r^* + ( 0.0104167 - 0.000526813 {\rm i} ) \Delta_r \Delta_r \Delta_a^* \nonumber \\
&+ ( 0.0208333 - 0.00105363 {\rm i} ) \Delta_a \Delta_r \Delta_r^* \nonumber \\
&+ (0.00466688 - 0.000263407 {\rm i}) \Delta_a \Delta_a \Delta_r^* \nonumber \\
&+ (0.00933375 - 0.000526813 {\rm i}) \Delta_a \Delta_r \Delta_a^* \nonumber \\
&+ (0.00233344 - 0.000258912 {\rm i}) \Delta_a \Delta_a \Delta_a^*, \nonumber \\
\psi_{b2}^{*(0)(3)(0)} =&- (0.00477255 + 0.0160286 {\rm i}) B_{1v}^2 \Delta_a^* \nonumber \\
&- (0.00571903 + 0.0354791 {\rm i} ) B_{1v}^2 \Delta_r^* \nonumber \\
&- (0.0761518 + 0.061066 {\rm i}) B_{2v}^2 \Delta_a^* \nonumber \\
&+ (0.167568 + 0.0545957 {\rm i}) B_{2v}^2 \Delta_r^* \nonumber \\
&- ( 0.00571904 + 0.0129802 {\rm i} ) B_{1v} B_{2v} \Delta_a^* \nonumber \\
&+ ( 0.0114381 - 0.0191166 {\rm i} ) B_{1v} B_{2v} \Delta_r^* \nonumber \\
&- \frac{1}{8} B_{ri}^2 \Delta_a^* + \frac{1}{4} B_{ri}^2 \Delta_r^* - \frac{8 + 8i \pi - 3 \pi^2}{64 \pi^2} B_{ai}^2 \Delta_r^* \nonumber \\
&+ (\frac{7}{96} - \frac{\rm i}{8 \pi}) B_{ai} B_{ri} \Delta_a^* - \frac{\pi - {\rm i}}{4 \pi} B_{ai} B_{ri} \Delta_r^* \nonumber \\
&+ (-0.0171049 + 0.00312451 {\rm i}) B_{ai}^2 \Delta_a^*  \nonumber \\
&- \frac{1}{48} \Delta_r^* \Delta_r^* \Delta_r - (0.0208333 + 0.00105363 {\rm i}) \Delta_a^* \Delta_r^* \Delta_r \nonumber \\
&- (0.0104167 + 0.000526813 {\rm i}) \Delta_r^* \Delta_r^* \Delta_a \nonumber \\
&+ (0.00466688 + 0.000263407 {\rm i}) \Delta_a^* \Delta_a^* \Delta_r \nonumber \\
&+ (0.00933375 + 0.000526813 {\rm i}) \Delta_a^* \Delta_r^* \Delta_a \nonumber \\
&- (0.00233344 + 0.000258012 {\rm i}) \Delta_a^* \Delta_a^* \Delta_a, \nonumber \\
\psi_{b1}^{*(0)(3)(0)} =& (0.0761517 - 0.0610661 {\rm i}) B_{1v}^2 \Delta_a^* \nonumber \\
&+ (0.167568 - 0.0545957 {\rm i}) B_{1v}^2 \Delta_r^* \nonumber \\
&+ (0.00477255 - 0.0160286 {\rm i}) B_{2v}^2 \Delta_a^* \nonumber \\
&+ ( -0.00571908 + 0.0354791 {\rm i} ) B_{2v}^2 \Delta_r^* \nonumber \\
&+ ( 0.00571904 - 0.0129802 {\rm i} ) B_{1v} B_{2v} \Delta_a^* \nonumber \\
&+ (0.0114381 + 0.0191166 {\rm i}) B_{1v} B_{2v} \Delta_r^* \nonumber \\
&+ \frac{1}{8} B_{ri}^2 \Delta_a^* + \frac{1}{4} B_{ri}^2 \Delta_r^* - \frac{8 - 8 {\rm i} \pi - 3 \pi^2}{64 \pi^2} B_{ai}^2 \Delta_r^* \nonumber \\
&+ \left( \frac{7}{96} + \frac{\rm i}{8\pi} \right) B_{ai} B_{ri} \Delta_a^* + \frac{{\rm i} +\pi}{4\pi} B_{ai} B_{ri} \Delta_r^* \nonumber \\
&+ (0.0171049+0.00312451 {\rm i}) B_{ai}^2 \Delta_a^*  \nonumber \\
&+ \frac{1}{48} \Delta_r^* \Delta_r^* \Delta_r + ( 0.0104167 - 0.000526813 {\rm i} ) \Delta_r^* \Delta_r^* \Delta_a \nonumber \\
&+ (0.0208333 - 0.00105363i) \Delta_a^* \Delta_r^* \Delta_r \\
& + (0.00466688 - 0.000263407 {\rm i}) \Delta_a^* \Delta_a^* \Delta_r \nonumber \\
&+ (0.00933375 - 0.000526813 {\rm i}) \Delta_a^* \Delta_r^* \Delta_a \nonumber \\
& + (0.00233344 - 0.000258012 {\rm i}) \Delta_a^* \Delta_a^* \Delta_a.
\end{align}
Then, the leading quartic terms in EFT action are
\begin{align}
\mathcal L_{eff}^{(0)(4)(0)}= & - 0.173287 B_{rv}^2 \left( \Delta_a \Delta_r^* + \Delta_r \Delta_a^* \right) - 0.346573 B_{av} B_{rv} \Delta_r \Delta_r^* \nonumber \\
& + 0.090075 {\rm i} B_{av} B_{rv} \left( \Delta_a \Delta_r^* + \Delta_r \Delta_a^* \right) + 0.0900764 {\rm i} B_{rv}^2 \Delta_a \Delta_a^* \nonumber \\
&+ 0.0191166 {\rm i} B_{av}^2 \Delta_r \Delta_r^* - 0.0376026 B_{av}^2 \left( \Delta_a \Delta_r^* + \Delta_r \Delta_a^* \right)\nonumber \\
&- 0.0713795 B_{av} B_{rv} \Delta_a \Delta_a^* + 0.0160281{\rm i} B_{av}^2 \Delta_a \Delta_a^* \nonumber \\
&- \frac{1}{4} B_{ri}^2 \left( \Delta_a \Delta_r^* + \Delta_r \Delta_a^* \right) + \frac{1}{2} B_{ri} B_{ai} \Delta_r \Delta_r^* - \frac{\rm i}{4\pi} B_{ai}^2 \Delta_r \Delta_r^* \nonumber \\
& - \frac{\rm i}{4\pi} B_{ri} B_{ai} \left( \Delta_a \Delta_r^* + \Delta_r \Delta_a^* \right) - 0.0342099 B_{ai}^2 \left( \Delta_a \Delta_r^* + \Delta_r \Delta_a^* \right) \nonumber \\
&- 0.0729167 B_{ri} B_{ai} \Delta_a \Delta_a^* - 0.00312451 {\rm i} B_{ai}^2 \Delta_a \Delta_a^*  \nonumber \\
&- 0.0208333 \Delta_r \Delta_r^* \left( \Delta_a  \Delta_r^* + \Delta_r \Delta_a^* \right) + 0.000263406 {\rm i} \left[ \left(\Delta_a \Delta_r^*\right)^2 + \left(\Delta_a^*\Delta_r \right)^2 \right] \nonumber \\
&- 0.00466688 \Delta_a \Delta_a^* \left(\Delta_a \Delta_r^* + \Delta_r \Delta_a^* \right) + 0.00105363 {\rm i} \Delta_a \Delta_r \Delta_a^* \Delta_r^* \nonumber \\
&+ 0.000129006 {\rm i} \Delta_a \Delta_a \Delta_a^* \Delta_a^*. \label{Leff_040}
\end{align}

Combining the results \eqref{Leff_030}, \eqref{Leff_130} and \eqref{Leff_040}, we obtain the interaction part \eqref{L_int} as well as the non-Gaussian terms in \eqref{L_Delta}. Particularly, holographic results prove that the chemical shift symmetry and dynamical KMS symmetry are nicely satisfied.

\section*{Acknowledgements}


YB was supported by the National Natural Science Foundation of China (NSFC) under the grant No. 12375044. XG and ZL were partially supported by NSFC under Grant Nos 12375065 and 12005150.

\bibliographystyle{utphys}
\bibliography{reference}

\end{document}